  \documentclass[prc,twocolumn,tightenlines,amssymb,aps,nobibnotes,
  superscriptaddress,showpacs,balancelastpage,floatfix,nofootinbib]{revtex4-1}
\usepackage{graphicx,colordvi}
\usepackage{multirow}      
\usepackage{color}	
\usepackage[utf8]{inputenc}
\usepackage{amsmath}	
\expandafter\ifx\csname package@font\endcsname\relax\else
\expandafter\expandafter
\expandafter\usepackage
\expandafter{\csname package@font\endcsname}
\fi
\DeclareRobustCommand\substyle{\name@idx{document substyle}}
\DeclareRobustCommand\classoption{\name@idx{document class option}}
\DeclareRobustCommand\classname{\name@idx{document class}}
\def\name@idx#1#2{{\ttfamily#2}
\index{#2\space#1=\string\ttt{#2}\space#1}\index{#1>#2=\string\ttt{#2}}}

\newcommand{\beq}[0]{\begin{equation}}
\newcommand{\eeq}[0]{\end{equation}}
\newcommand{\bea}[0]{\begin{eqnarray}}
\newcommand{\eea}[0]{\end{eqnarray}}
\def\d{\hbox{d}}
\def\be{\begin{equation}}
\def\ee{\end{equation}}
\def\bea{\begin{eqnarray}}
\def\eea{\end{eqnarray}}
\def\l{\label}

\def\hahat{\hat{H}}

\def\hahat0{\hat{H}_0}

\def\cos{\hbox{cos}}

\def\sinh{\hbox{sinh}}

\def\exp{\hbox{exp}}

\def\sinh{\hbox{sinh}}

\def\vareps{\varepsilon}

\def\siml{\hbox{\kern.1em \lower.6ex \hbox{$\sim$} \kern-1.12em
          \raise.6ex \hbox{$<$} \kern.1em }}
\def\simg{\hbox{\kern.1em \lower.6ex \hbox{$\sim$} \kern-1.12em
          \raise.6ex \hbox{$>$} \kern.1em }}

\def\siml{\hbox{\kern.1em \lower.6ex \hbox{$\sim$} \kern-1.12em
 \raise.6ex \hbox{$<$} \kern.1em}}
\def\simg{\hbox{\kern.1em \lower.6ex \hbox{$\sim$} \kern-1.12em
 \raise.6ex \hbox{$>$} \kern.1em}}

\newcommand{\beqar}{\begin{eqnarray}}
\newcommand{\eeqar}[1]{\label{#1} \end{eqnarray}}

\begin{document}

\title{Semiclassical shell-structure  micro-macroscopic approach for the level
  density}
\author{A.G. Magner}
\email{Email: magner@kinr.kiev.ua}
\affiliation{\it  Institute for Nuclear Research, 03028 Kyiv, Ukraine} 
\author{A.I. Sanzhur}
\affiliation{\it  Institute for Nuclear Research, 03028 Kyiv, Ukraine} 
\author{S.N. Fedotkin}
\affiliation{\it  Institute for Nuclear Research, 03028 Kyiv, Ukraine}
\author{A.I. Levon}
\affiliation{\it  Institute for Nuclear Research, 03028 Kyiv, Ukraine}
\author{S. Shlomo}
\affiliation{\it Cyclotron Institute, Texas A\&M University,
  College Station, Texas 77843, USA} 
%
%\date{\today}
%
\bigskip
\date{March, 24th, 2021}
\bigskip

\begin{abstract}

  Level density $\rho(E,A)$ is derived for a 
 one-component nucleon 
  system with
  a given energy $E$ and particle number $A$
   within the mean-field semiclassical periodic-orbit theory 
  beyond the saddle-point method of the Fermi gas model.
  We obtain $~~\rho \propto I_\nu(S)/S^\nu$,~~ 
  with $I_\nu(S)$ being
  the
  modified Bessel function
  of the entropy $S$.
  Within the micro-macro-canonical approximation (MMA),
  for a small thermal excitation energy, $U$, with respect to
    rotational excitations, $E_{\rm rot}$, 
   one obtains $\nu=3/2$ for $\rho(E,A)$. 
   In the case  of 
   excitation energy $U$ larger than $E_{\rm rot}$ but smaller than
   the neutron separation energy, one finds a larger value of $\nu=5/2$.
   A role of the
   fixed spin variables for rotating nuclei is discussed.
   The MMA level density $\rho$ 
  reaches 
  the well-known 
   grand-canonical ensemble limit (Fermi gas asymptotic)
  for large 
  $S$ related to large excitation energies,
  and also reaches the finite micro-canonical limit for small combinatorial entropy $S$
  at low excitation energies 
      (the constant ``temperature'' model).
  Fitting the $\rho(E,A)$ of the MMA 
  to the 
  experimental data for low excitation energies,
  taking into account shell and, qualitatively, pairing effects, 
  one obtains for
   the
    inverse level
    density parameter $K$ a value 
    which differs essentially from  that parameter
    derived from data on 
    neutron resonances.

\end{abstract}
\maketitle

\noindent
%PACS numbers: 21.10. Ev, 21.60. Cs, 24.10 Pa

%%%%%%%%%%%%%%%%%%%%%%%%%%%%%%%%%%%%%%%%%%%%%%%%%%%%%%%%%%%%%%%%%%%%%%%%
\section{Introduction}
\label{sec-introd}

Many properties of heavy nuclei can be 
described
in terms of the statistical level density
\cite{Be36,Er60,GC65,BM67,St72,LLv5,Ig83,So90,Sh92,Ra97,Ig98,AB00,AB03,EB09,
 Ba09,Ch10, AB16,ZS16,HJ16,KS18,ZK18,Ze19,KS20}.
 A well-known old example is 
the description
of neutron resonances  
using the level density.
Usually, the level density 
$\rho(E,A)$, where $E$ and $A$ 
are the energy and
  nucleon number,
respectively,
is given by
the inverse Laplace transformation of the partition function
$\mathcal{Z}(\beta,\alpha)$.
Within the grand
canonical ensemble 
the standard saddle-point method (SPM) is used
for integration
over all variables,
including 
$\beta$, which is related to the total energy $E$ \cite{Er60,BM67}.
This method assumes large
excitation energies $U$, so  that the temperature $T$ is
related to a  well-determined
saddle point
in the integration variable
$\beta$ for a finite Fermi 
    system of large particle numbers. 
However,  data from
many experiments
 for energy levels and spins also exist for regions 
of 
low excitation energy $U$,
where  such a saddle point does not exist. 
For presentation of experimental data on nuclear spectra, the cumulative
level-density distribution 
-- cumulative number of quantum levels
below the excitation energy $U$ 
--
is conveniently often used 
for statistical analysis 
\cite{Ze96,Go11,ML18} of
the experimental data on collective excitations
\cite{Le94,ML18,Le19,Le19a}. 
    For calculations of 
this cumulative level density, 
one has to
integrate the level density
over a large interval of the excitation energy $U$. 
    This interval extends
   from small
 values of  $U$, where
there is no 
 thermodynamic equilibrium
(and no meaning 
 to the temperature), to large
 values of $U$, where the standard grand
canonical ensemble 
can be successfully
applied in terms of the temperature $T$ in a finite Fermi system.
 Therefore, 
to simplify the calculations of the level density, $\rho(E,A)$, 
 we 
will, in the following, carry out the integration over the Lagrange multiplier
$\beta$ in the inverse Laplace
  transformation of the partition function
$\mathcal{Z}(\beta,\alpha)$  
  more 
accurately  beyond the SPM \cite{KM79,MK79,PLB}. 
 However, for a nuclear system with large particle number $A$ one can apply the SPM
    for the variable
$\alpha$, related to $A$.
The case of
  neutron-proton asymmetry of the Fermi system will be worked out
  separately.
Thus, for remaining
integration over $\beta$ we shall use approximately the
micro-canonical ensemble 
which does not assume
a temperature and
an existence of thermodynamic equilibrium.
Notice that there are other methods to overcome divergence of
the full SPM for low
excitation-energy limit $U \rightarrow 0$; see 
Refs.~\cite{JB75,BJ76,PG07,ZS16,ZK18}.
The 
well-known method 
suggested
in Ref.~\cite{JB75} 
is applied successfully for the partition function of
    the extended Thomas-Fermi (ETF) theory at
    finite temperature to obtain the smooth level density and
    free energy; see also Refs.~\cite{BJ76}
and \cite{BB03}, and references therein.

For formulation of the 
unified microscopic 
canonical 
and macroscopic grand-canonical 
approximation
 (MMA)   
to the level density,
we will find
a simple analytical
approximation for the
level density $\rho$ 
which satisfies
the two well-known limits. 
One of them is the 
Fermi gas asymptotote, 
$\rho \propto \exp(S)$, with the entropy
$S$,  
 for 
large entropy $S$.  Another limit is 
the combinatorics expansion in powers of $S$ for a small entropy $S$
or excitation energy $U$,
always at large particle numbers $A$; see Refs.\ \cite{St58,Er60,Ig72,Ig83}.
The empiric formula,
    $\rho\propto \exp[(U-E_0)/T]$ with free parameters $E_0$,  $T$, and
  a  preexponent factor,
was suggested for the description of the excited low energy states (LESs) in
Ref.\ \cite{GC65}.
Later, this formula was 
 named 
the constant
``temperature'' model (CTM) 
where the ``temperature'' is considered 
an ``effective temperature'' 
related to the excitation energy
 (with no direct physical meaning
of temperature for LESs); see also Ref.~\cite{ZK18,Ze19}. 
We will show below that 
the MMA
has the same power 
expansions as the CTM for LES at small excitation energies
$U$. 
We will 
also 
show that, within
the MMA, the transition between
these two limits 
is sufficiently
rapid, when
considered over the dimensionless entropy variable $S$. 
  Therefore, our aim is  to derive approximately a simple
statistically averaged analytical expression for the level density $\rho(S)$
with the correct two 
    limits, mentioned above, for small and large values of $S$.

Such an MMA for  the
level density $\rho$ was 
suggested in Refs.~\cite{KM79,MK79} 
in terms of the modified Bessel function
of the entropy variable in the case of
    small
    excitation energy $U$ as compared to the rotational energy
    $E_{\rm rot}$. 
The so-called a ``classical rotation''
of the spherical or axially symmetric nucleus was considered  
alignment of nucleons
along the symmetry axis on the basis of the periodic orbit theory with
a fixed angular momentum and its projection (see Ref.~\cite{MK78}),
in contrast to the collective rotation around the
perpendicular axis
\cite{MG17,GM21}.
The yrast
line was defined 
to be at zero
  excitation energy for a given angular
  momentum within the cranking model 
 \cite{In54,RS80}. 
One of the important characteristics
 of 
the yrast line
is the
moment of inertia (MI).
The Strutinsky shell-correction method (SCM) 
\cite{St67,BD72},
extended by Pashkevich and 
Frauendorf 
\cite{PF75} to the 
description of 
nuclear rotational bands,
was applied \cite{KM79,MK79} for studying the shell effects in the MI
near
the yrast line.

For a deeper understanding of the 
correspondence between the classical and the quantum approach, especially
their applications
to 
high-spin
physics, it is worthwhile to analyze the shell 
effects in the level density $\rho$
(see Refs.~\cite{Ig83,So90}), 
 in particular, in the entropy $S$ and MI, within the semiclassical
periodic-orbit (PO) theory (POT) 
\cite{Gu71,BB72,Gu90,SM76,SM77,BB03,MY11,MG17,MK78,GM21}.
This theory, based on the semiclassical time-dependent propagator,
 enables determining
    the total level-density, energy,
free-energy, and grand
canonical ensemble 
potential in terms of the smooth ETF term and 
PO-shell corrections \cite{SM76,SM77,BB03,MY11,MG17,KM79,MK79}.

We 
will extend the MMA
approach \cite{KM79}, 
 in order to consider the 
shell effects in
the yrast line as a minimum of the
nuclear level density (minimum excitation energy),
for the description of shell and collective effects in
terms of the level density 
itself for
larger 
excitation energies $U$.
The level
density parameter $a$
is one of the key quantities
under 
intensive experimental and theoretical 
investigations; see, e.g.,
Refs.~\cite{Be36,Er60,GC65,BM67,St72,Ig83,Sh92,So90,EB09,KS20}.
 Mean values of $a$ are largely
proportional to
the particle number $A$.
The inverse level density parameter
$K=A/a$ is conveniently introduced
to exclude 
a basic mean $A$ dependence in $a$.
Smooth properties of $K$ 
as function of the nucleon number $A$
have been studied
within the framework of the
self-consistent 
ETF approach \cite{Sh92,KS18}.
However, for instance,
shell 
effects in the statistical
level density 
are still 
an attractive subject. 
This is due to the major shell
effects in the distribution of single-particle (s.p.) 
states near
the Fermi surface
within the mean-field approach.
The nuclear shell
effects influence
the statistical level density
of a heavy nucleus,  which is especially important near magic numbers,
see Refs. \cite{Ig83,So90} and references therein.
In the present study,  for simplicity, we  shall first work out 
the derivations of the level
density $\rho(E,A)$ for a 
one-component nucleon system, taking into account the shell,
    rotational  and, qualitatively,  pairing effects.
This work is concentrated on LESs of
nuclear excitation-energy spectra below
the neutron resonances.

The paper is organized as the following.
The level density $\rho(E,A)$ is derived
within the MMA 
by using the POT in Sec. \ref{sec-levden}.
We extend the MMA to 
large excitation energies $U$,
up to about the neutron separation energy,
taking essentially into account the shell effects.
Several analytical approximations, 
    in particular the spin dependence of the
level density  are presented in Sec. \ref{sec-MMAas}.
Illustrations of the MMA for the level
density $\rho(E,A)$ and inverse level density parameter $K$
versus experimental
 data, 
    discussed
for
typical heavy nuclei,  are given in Sec. \ref{sec-disc}.
 Our conclusions are presented in Sec. \ref{sec-concl}.
 The semiclassical POT is described in  Appendix \ref{appA}.  
The level density, $\rho(E,A,M)$, derived
by accounting for the 
rotational excitations
with the fixed projection of the angular momentum
$M$ and spin 
$I$ of nuclei
in the case of spherically symmetric or
 axially 
 symmetric mean fields, is given in Appendix \ref{appB}.
 The full SPM level density
 derivations generalized by shell effects are 
  described in Appendix 
 \ref{appC}.

 \section{Microscopic-macroscopic approach }
 \l{sec-levden}
 For 
a statistical
description of level density of a nucleus in
  terms of the  conservation variables,
the total energy, $E$, and nucleon number,  $A$, 
one
can begin with
the micro-canonical expression for the level density,
\begin{equation}\label{dendef1}
\rho(E,A)=
\sum\limits_i\!\delta(E-E_i)~\delta(A-A_i)
\equiv
\int \frac{\d \beta \d \alpha}{(2\pi i)^2}~e^{S},
\end{equation}
where $E_i$ and $A_i$ represent the system spectrum, and
$S=\ln \mathcal{Z}(\beta,\alpha)
+\beta E -\alpha A~$
is the
entropy.
Using the mean field approximation for the partition function
$\mathcal{Z}(\beta,\alpha)$, one finds \cite{BM67}  
\bea\l{parfunF}
&\ln \mathcal{Z}= 
\sum\limits_{i}\ln\left[1 +
\exp\left(\alpha - 
   \beta \varepsilon_i\right)\right]\nonumber\\
&\approx \int\limits_0^{\infty}\d \varepsilon~g(\varepsilon)
\ln\left[1+
\exp\left(\alpha -
    \beta\varepsilon\right)\right]~,
\eea
where $\varepsilon_i$  are the
s.p. energies 
of the quantum states in  the mean field. In the transformation
from the sum to an
integral, we introduced  the s.p. level density $g(\varepsilon)$ %of the SCM
as a sum of
the smooth, $\tilde{g}(\varepsilon)$, 
and oscillating shell,
$\delta g(\varepsilon)$, 
components, using the SCM (see Refs.~\cite{St67,BD72}):
\be\l{spden}
g(\varepsilon)\cong \tilde{g}(\varepsilon)+
\delta g(\varepsilon)~.
\ee
Within the semiclassical POT \cite{SM76,BB03}, the smooth and oscillating parts
of the s.p. level  density, $g(\varepsilon)$, can
be approximated, with good accuracy, by the sum of
the ETF level density,
$\tilde{g} \approx g^{}_{\rm ETF}$,
and the semiclassical PO contribution,
$\delta g(\varepsilon)\approx  \delta g_{\rm scl}$,
Eq.~(\ref{goscsem}).
 In integrating over $\alpha$ in Eq.~(\ref{dendef1}) for a  given $\beta$ by the
standard SPM, 
we
use the expansion 
    for the entropy
$S(\beta,\alpha)$ near the saddle point %(SP)
$\alpha=\alpha^\ast$
as
\be\l{Sexp}
S(\beta,\alpha)=S(\beta,\alpha^\ast)
+\frac12 \left(\frac{\partial^2 S}{\partial \alpha^2}\right)^\ast
\left(\alpha-\alpha^\ast\right)^2+\ldots~.
\ee
The first-order term of this expansion disappears because
the Lagrange multiplier, $\alpha^\ast$,
is defined by the 
saddle-point condition
\begin{equation}\label{Seqsd}
\left(\frac{\partial S}{\partial \alpha}\right)^\ast\equiv
\left(\frac{\partial \ln Z}{\partial \alpha}\right)^\ast-A=0~.
\end{equation}

Introducing, for convenience, the potential
$\Omega=-\mbox{ln}\mathcal{Z}/\beta$, one can use its SCM decomposition in terms of the smooth part and shell
corrections for the level density
$g$, see Eq.~(\ref{spden})  and Ref.~\cite{KM79}, through the
partition function, $ \ln \mathcal{Z}$ (Eq.~(\ref{parfunF})):
  \be\l{SCMpotF}
  \Omega\left(\beta,\lambda\right) 
   \cong 
  ~\tilde{\Omega}\left(\beta,\lambda\right) 
  +
  \delta \Omega\left(\beta,\lambda\right)~.
  \ee
   Here, $\tilde{\Omega}\approx \Omega^{}_{\rm ETF}$ is the smooth
  ETF component \cite{KM79,KS20},  
\be\l{TFpotF}
\tilde{\Omega}\left(\beta,\lambda\right)
=\tilde{E} 
-\lambda A
-\frac{\pi^2}{6\beta^2}\tilde{g}(\lambda)~,
\ee
where $\tilde{E}\approx E^{}_{\rm ETF}$ is the
nuclear ETF energy (or the liquid-drop
energy). 
For a given $\beta$, the chemical  potential, 
$\lambda=\alpha^\ast/\beta$,
is a function of the particle number $A$, 
 according to Eq.~(\ref{Seqsd}), and
$\lambda \approx \tilde{\lambda}$ is  approximately equal to
the SCM smooth chemical potential.
With the help of the POT
\cite{SM76,SM77,BB03}, one obtains \cite{KM79} for the oscillating (shell)
component, $\delta \Omega$, in Eq.~(\ref{SCMpotF}),
\bea\l{potoscparF}
&\delta \Omega= -\beta^{-1} 
\int\limits_0^\infty\d\varepsilon~
\delta g(\varepsilon)~
\ln\left\{1+\exp\left[\beta\left(\lambda-
  \varepsilon\right)\right]\right\}\nonumber\\
&\cong
\delta \Omega_{\rm scl}\left(\beta,\lambda\right)
=\delta F_{\rm scl}~.
\eea
For the
semiclassical free-energy shell correction, $\delta F_{\rm scl}$
(see Appendix \ref{appA}), we incorporate 
the POT expression: 
\be\l{FESCF}
\delta F_{\rm scl} \cong \sum^{}_{\rm PO} F_{\rm PO}~,
\ee
 where,
\be\l{dFESCF}
F_{\rm PO}= E_{\rm PO}~
\frac{x^{}_{\rm PO}}{
  \sinh\left(x^{}_{\rm PO}\right)}~,\quad x^{}_{\rm PO}=
\frac{\pi t^{}_{\rm PO}}{\hbar \beta}~,
\ee
and
\be\l{dEPO0F}
E_{\rm PO}=\frac{\hbar^2}{t_{\rm PO}^2}\,
g^{}_{\rm PO}(\lambda)~.
\ee
 Here, $t^{}_{\rm PO} = k~t^{k=1}_{\rm PO}(\lambda)$ 
is the period 
of particle motion
along the PO (taking into
account its repetition, or period number $k$),
and
$t^{k=1}_{\rm PO}$ is the
period of the particle motion along the
primitive 
($k=1$) PO.
The period $t^{}_{\rm PO}$ (and $t^{k=1}_{\rm PO}$) and
the partial oscillating level density component, $g^{}_{\rm PO}$,
given by Eq.~(\ref{goscPO}),
are taken at the chemical potential $\varepsilon=\lambda$;
see also Eqs.~(\ref{goscsem}) and (\ref{goscPO})
for the semiclassical s.p. level-density shell correction
$\delta g_{\rm scl}(\varepsilon)$
(see Refs.~\cite{SM76,BB03}). 
    Notice that equivalence of the variations
    of the 
    grand-canonical- and canonical- ensemble potentials,  
    Eq.~(\ref{potoscparF}), is valid
    approximately in the 
 corresponding
     variables, 
        for large particle  numbers $A$.
      This equivalence has to be 
    valid in
the semiclassical POT.

Expanding, then, $x^{}_{\rm PO}/\sinh(x^{}_{\rm PO})$,
Eq.~(\ref{dFESCF}), in the shell correction $\delta \Omega$
[Eqs.~(\ref{potoscparF}) and (\ref{dFESCF})] 
in powers of  $1/\beta^2$
up to the quadratic terms, $\propto 1/\beta^2$, one obtains
\be\l{OmadF}
\Omega \approx E_0-\lambda A-\frac{a}{\beta^2}~.
\ee
Here $E_0$  is the ground state energy, $E_0=\tilde{E}+\delta E$,
and $\delta E$ is the energy shell correction of a cold nucleus,
$\delta E \approx \delta E_{\rm scl}$, 
 Eq.~(\ref{escscl}).
In Eq.~(\ref{OmadF}),
$a$
is the level density parameter $a$, 
\be\l{denpar}
a=\tilde{a}+\delta a~,
\ee
where $\tilde{a}\approx a^{}_{\rm ETF}$  and $\delta a$ are the ETF and
the shell correction components,
\be\l{daF}
\tilde{a} 
\approx \frac{\pi^2}{6} g^{}_{\rm \tt{ETF}}(\lambda), \quad
\delta a=\frac{\pi^2}{6}\delta g_{\rm scl}(\lambda)~.
\ee
 Note that for the ETF components one commonly accounts for self-consistency using
Skyrme interactions, see Refs.~\cite{BG85,BB03,AS05,KS18,KS20,PLB}.
For the semiclassical POT level density, $\delta g_{\rm scl}(\lambda)$,
 one employs the method of Eqs.~(\ref{goscsem})
and (\ref{goscPO}), 
see Refs.~\cite{BB72,SM76,SM77,BB03,MY11,MG17}.
Note that in the grand canonical ensemble, the level density
parameter $a$, Eqs.~(\ref{denpar}) with (\ref{daF}),
    is function of the chemical potential $\lambda$.
We may  include, generally speaking, the collective (rotational) component
into $E_0$;
see Sec.~\ref{subsec-I} and Appendix \ref{appB}.

Substituting Eq.~(\ref{Sexp}) into Eq.~(\ref{dendef1}),
and
taking the error integral over $\alpha $ in the extended infinite
limits including the saddle point $\alpha^\ast$, one obtains
\bea\l{rhoE1F}
&\rho(E,A) \approx 
\frac{1}{2\pi i~\sqrt{2\pi}}
\int \d \beta~\beta^{1/2}
\mathcal{J}^{-1/2}\nonumber\\
&\times \exp\left(\beta U + a/\beta\right)~,
\eea
where $U=E-E_0$ is the excitation energy,  and $a$ is the level density parameter,
given by Eqs.~(\ref{denpar}) and (\ref{daF}). In 
equation (\ref{rhoE1F}),
$\mathcal{J}$ is the one-dimensional 
 Jacobian determinant 
 [$c$ number, $\mathcal{J}(\lambda)$] taken 
 at the saddle point over $\alpha$ at $\alpha=\alpha^\ast=\lambda \beta$,
 Eq.~
(\ref{Seqsd}):
\bea\l{Jac1F}
&\mathcal{J}\equiv\beta \left(\frac{\partial^2 S}{\partial \alpha^2}\right)^\ast
\equiv \beta \left(\frac{\partial^2 \ln Z}{\partial \alpha^2}\right)^\ast\nonumber\\
&=-\left(\frac{\partial^2\Omega}{\partial \lambda^2}\right)^\ast
 \cong \tilde{\mathcal{J}}+\delta \mathcal{J}~.
\eea
The asterisks mean the saddle point 
for integration over $\alpha$ for any $\beta$
(here and in the following 
we omit the superscript  asterisk in $\mathcal{J}$).
Differentiating 
 the potential $\Omega$, Eq.~(\ref{SCMpotF}),  over $\lambda$
within the  grand-canonical ensemble  we 
 obtain for the smooth part of the Jacobian
$\tilde{\mathcal{J}}=
-\left(\partial^2\Omega^{}_{\rm \tt{ETF}}/\partial\lambda^2\right)^\ast
 \approx g^{}_{\rm \tt{ETF}}(\lambda)$.
 We note that,
for not too large thermal excitations,
the main contribution from the oscillating potential component
$\delta \Omega$ as function of $\lambda$ is coming from the
differentiation of the  sine  function in the
PO energy shell correction factor $E_{\rm PO}$, Eq.~(\ref{dEPO0F}), through
the PO action phase
$\mathcal{S}_{\rm PO}(\lambda)/\hbar$ of
the PO level density  
component $g^{}_{\rm PO}(\lambda)$, Eq.~(\ref{goscPO}).
 The temperatures $T=1/\beta^\ast$,
 when 
 the saddle point $\beta=\beta^\ast$ exists, are assumed to be 
much smaller than the chemical
potential $\lambda$. The reason is that
  for large particle numbers $A$ the
 semiclassical large parameter,
$\sim\mathcal{S}_{\rm PO}/\hbar \sim A^{1/3}$,  appears.
This leads to 
a dominating contribution, much larger than that coming
from differentiation of other terms, 
the $\beta$-dependent function
$x^{}_{\rm PO}(\beta)$, and 
the PO period $t^{}_{\rm PO}(\lambda)$.
 Using Eqs.~(\ref{potoscparF}), 
(\ref{d2Edl2}), 
    and (\ref{d2g}),  one 
 approximately  obtains
for the oscillating Jacobian part $\delta \mathcal{J}(\lambda)$,
Eq.~(\ref{Jac1F}),  the  expression:
\be\l{dJ}
\delta \mathcal{J} \approx\sum^{}_{\rm PO}g^{}_{\rm PO}\frac{x^{}_{\rm PO}}{
  \sinh\left(x^{}_{\rm PO}\right)}~,
\ee
where $x^{}_{\rm PO}(\beta,\lambda)$ [through $t^{}_{\rm PO}(\lambda)$]
is the dimensionless quantity, Eq.~(\ref{dFESCF}),
proportional
to $1/\beta$. 
The total Jacobian $\mathcal{J}(\lambda)$ as function of $\lambda$
can be presented as  
\be\l{Jac2}
\mathcal{J} \cong \tilde{\mathcal{J}}\left(1+\delta \mathcal{J}/\tilde{\mathcal{J}}\right)
=g(\lambda)\left(1+\xi\right),
\ee
where $\xi(\beta,\lambda)$ is defined by [see also Eqs.~(\ref{Jac1F})
and (\ref{OmadF})]
\be\l{xipar}
\xi=\frac{a^{\prime\prime}(\lambda)}{\beta^2g(\lambda)}\approx
\sum^{}_{\rm PO}\frac{g^{}_{\rm PO}(\lambda)}{g(\lambda)}
\left(\frac{x^{}_{\rm PO}}{\sinh\left(x^{}_{\rm PO}\right)}-1\right).
\ee
 This approximation 
   was derived
     for the case when a smooth (E)TF part can be neglected.
Notice, that the 
 rotational 
excitations can be included
into the ETF part and shell corrections of the potential $\Omega$; see Sec.~\ref{subsec-I}
and Appendix \ref{appB}.
In this case, Eq.~(\ref{Jac2}) will be 
similar but
with more complicate expressions for the two-dimensional Jacobian
$\tilde{\mathcal{J}}$, especially for its shell component  $\delta \mathcal{J}$.

Substituting now  
$\lambda$, found from Eq.~(\ref{Seqsd}),
for a given particle
number $A$, one can
obtain relatively small thermal and shell corrections to the smooth chemical potential 
    in 
$\lambda(A)$ of the SCM \cite{BD72}. 
    For simplicity, neglecting these correction terms 
    for large particle numbers, $A^{1/3}\gg 1$,
one can consider
$\lambda$
as a constant related to that of the particle number density of
nuclear  matter; see Sec. 2.3 of Ref.~\cite{BM67}.
    Therefore,  
$\lambda$ is
independent of the particle number $A$ for large values of $A$.

\section{MMA analytical expressions} 
\l{sec-MMAas}

 In 
linear approximation in $1/\beta^2$, one finds from Eq.~(\ref{xipar}) for $\xi$
and
Eq.~(\ref{dFESCF}) for $x^{}_{\rm PO}$
\be\l{xiLIN}
\xi =\frac{\overline{\xi}}{\beta^2}\approx
-\frac{\pi^2}{6\hbar^2\beta^2}
\sum^{}_{\rm PO}t^{2}_{\rm PO}\frac{g^{}_{\rm PO}(\lambda)}{g(\lambda)}~,
  \ee 
  where
\be\l{xib}
\overline{\xi}=\frac{a^{\prime\prime}(\lambda)}{g(\lambda)}\approx
-\frac{\pi^2}{6\hbar^2}
\sum^{}_{\rm PO}t^{2}_{\rm PO}\frac{g^{}_{\rm PO}(\lambda)}{g(\lambda)} 
\approx -\frac{2\pi^4}{3 D_{\rm sh}^2}
\frac{\delta g(\lambda)}{g(\lambda)}~;
  \ee
 see also Eq.~(\ref{xipar}).  In Eq.~(\ref{xib}),
  $D_{\rm sh}\approx \lambda/A^{1/3}$ is the distance between major shells;
 see Eq.~(\ref{periode}).
 For convenience,
introducing 
the 
 dimensionless energy shell
  correction,
$\mathcal{E}_{\rm sh}$, 
  in units of   
      the smooth ETF energy per particle, $E_{\rm \tt{ETF}}/A$,
      one can present
Eq.~(\ref{xib}) as
\be\l{xibdE}
\overline{\xi}
\approx
\frac{4\pi^6 A^{1/3}\mathcal{E}_{\rm sh}}{3\lambda^2}~, \quad
\mathcal{E}_{\rm sh}=-\frac{\delta E}{E_{\rm \tt{ETF}}}~A~.
  \ee
   In the applications below we will use
  $\overline{\xi}>0$ and $\mathcal{E}_{\rm sh}>0$ 
   if $\delta E<0$. 
    The smooth ETF energy $E_{\rm \tt{ETF}}$
   in Eq.~(\ref{xibdE})
    [see Eq.~(\ref{TFE0})]
    can be 
    approximated as 
      $E_{\rm \tt{ETF}}\approx \tilde{g}(\lambda)\lambda^2/2 $. 
  The energy shell correction, $\delta E$,
  was approximated,
  for a major shell structure,  with
  the semiclassical POT accuracy 
  (see Eqs.~(\ref{escscl}) and (\ref{dEPO0F}),
  and Refs.~\cite{SM76,SM77,BB03,MY11}) by,
  \be\l{dedg}
  \delta E \approx \delta E_{\rm scl}\approx
  \left(\frac{D_{\rm sh}}{2 \pi}\right)^2~ \delta g^{}_{\rm scl}(\lambda)~.
  \ee

  The correction $\propto 1/\beta^4$ of the expansion of the Jacobian (\ref{Jac2})
 in $1/\beta$ through the oscillating part $\delta \mathcal{J}$, Eq.~(\ref{dJ}),
 is 
 relatively small
 for $\beta $ which, at the saddle point 
 values $T=1/\beta^\ast$, is related to the
  chemical potential $\lambda$
  as $T \ll \lambda$. The high order,
  $\propto 1/\beta^4$, term of this expansion  can be
  neglected under the following condition:
 \be\l{condU}
\frac{1}{\tilde{g}}\siml U\ll
\sqrt{\frac{90}{7}}\frac{A^{1/3}\lambda^2}{2\pi^4 K}~.% \sim \lambda~.
\ee
 Using typical values for parameters $\lambda=40$ MeV, $A=200$,
and $K\approx 10$ MeV, $1/\tilde{g}\approx 0.1-0.2$ MeV;
see Ref.~\cite{KS18}; we may  approximately evaluate very
right-hand-side of Eq.~(\ref{condU}) as  20 MeV.
For simplicity,
           small shell and temperature corrections  to $\lambda(A)$
           from the conservation equation (\ref{Seqsd})
           are neglected by
       using linear shell effects of the leading order \cite{BD72} and constant particle
       number density of 
       nuclear matter, $\rho_0^{}$. Taking
       $\rho^{}_0=2k_F^{3}/3\pi^2=0.16$ fm$^{-3}$, one finds about constant
       $\lambda=\hbar^2k^{2}_F/2\mu\approx 40$ MeV, where $\mu$ is the nucleon mass.
       In the derivations of the condition (\ref{condU}),
       we used the POT 
       distance between major shells,
$D_{\rm sh}$, Eq.~(\ref{periode}).
Evaluation of the upper limit for the
excitation energy at the saddle point 
$\beta=\beta^\ast=1/T$
is justified  because this upper
limit is always so large that this point 
does  certainly exist.
  Therefore, for consistence, one can neglect the
  quadratic, $1/\beta^2$ (temperature $T^2$), corrections to the Fermi
  energy $\varepsilon^{}_{F}$ in the chemical potential,
  $\lambda\approx \varepsilon^{}_{F}$, for large particle numbers $A$.

  Under 
  the condition of Eq.~(\ref{condU}),
one can obtain simple analytical expressions for the level density
$\rho(E,A)$ from the integral representation (\ref{rhoE1F}), because
the
Jacobian factor $\mathcal{J}^{-1/2}$ in its integrand can be 
simplified
by expanding in small values of $\xi$ or of $1/\xi$
[see Eq.~(\ref{xiLIN})].
 Notice that one has
two terms in the Jacobian $\mathcal{J}$, Eq.~(\ref{Jac2}). One of them is
independent of the integration variable $\beta$ and 
 the other one is
proportional to
$1/\beta^2$. These two terms are connected 
 to those of
the potential $\Omega$, Eq.~(\ref{OmadF}), by the
inverse Laplace transformation (\ref{dendef1}) of the partition
function
(\ref{parfunF}) and the corresponding direct operation transformation.
    Expanding the
square root $\mathcal{J}^{-1/2}$ in the integrand of 
the integral representation (\ref{rhoE1F}),
for small and large $\xi$ at linear order in $\xi$ and $1/\xi$, respectively,
one arrives at two different approximations marked below as (i) and (ii) cases, respectively.
At each finite order of these expansions, one can  exactly take
the inverse Laplace transformation.
Convergence of the corresponding corrections to the level density,
Eq.~(\ref{rhoE1F}),
after applying the inverse
transformation, Eq.~(\ref{Laplace}), will be considered
in the next 
subsections.

%%%% FIG. 1:  %%%%%%%
\vspace{0.2cm}
\begin{figure*}
      \includegraphics[width=11.5cm]{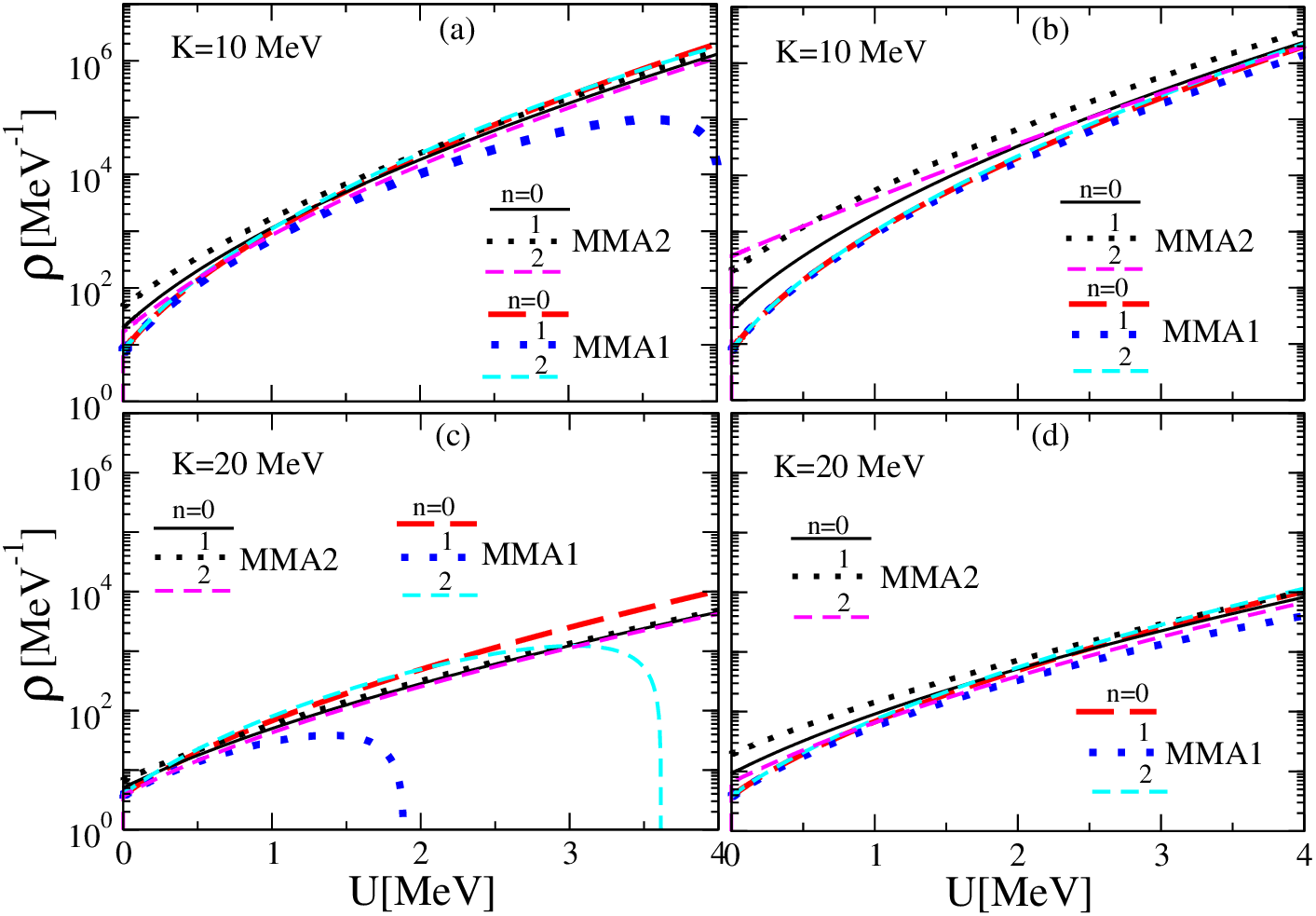}

\vspace{-0.2cm}
\caption{{\small 
    MMA level density $\rho$ [Eq.~(\ref{denbes}) in units of MeV$^{-1}$]
    as function of the excitation energy $U$
    (in units of MeV) at the inverse level density parameter $K=10$ MeV (a,b),
    and at $20$ MeV (c,d) 
    for the relative energy shell corrections
    $\mathcal{E}_{\rm sh}=1.7$ (a,c) and $0.6$
    (b,d)  
     values. 
     The black solid ($n=0$) and dotted ($n=1$) lines are of 
          MMA2,
     without [Eqs.~(\ref{rho52})]
     and with [Eq.~(\ref{rhoGEN52})] 
     the second term, respectively.
    The magenta dashed line
    ($n=2$) [numerical, Eq.~(\ref{rhoE1F})] with the 
  next leading correction term 
    presents good convergence  to the
    MMA2 results owing to the expansion of the Jacobian factor, $\mathcal{J}^{-1/2}$
   [see Eq.~(\ref{Jac2}) for the Jacobian $\mathcal{J}$], in the integrand
    of Eq.~(\ref{rhoE1F}),
    over
    $1/\xi$
 (see text).
    The heavy dashed red line ($n=0$) and blue dotted line ($n=1$), and the
    dashed cyan line
($n=2$) [see Eqs.~(\ref{rho32})  (MMA1)
and (\ref{rhoGEN32}), and (\ref{rhoE1F}),
    respectively],
    show the convergence to the MMA1 results due 
    to the expansion of this Jacobian $\mathcal{J}$, 
     over
    $\xi$.
    The particle number 
    $A=200$ was used.    
   }}
\label{fig1}
\end{figure*}

\subsection{(i) Small shell effects}
\l{subsec-32}

Using Eq.~(\ref{Jac2}), one can write for small $\xi$, 
Eq.~(\ref{xiLIN}),
\be\l{Jac3}
\frac{1}{\mathcal{J}^{1/2}}=\frac{1}{\sqrt{g(\lambda)\left(1+\xi\right)}}\approx
\frac{1}{\sqrt{g(\lambda)}}\left(1-\frac{\overline{\xi}}{2\beta^2}\right)~. 
\ee
Substituting this expression for the Jacobian factor, $\mathcal{J}^{-1/2}$, into
Eq.~(\ref{rhoE1F}) one obtains two terms, which 
are related to
those of the last equation in
(\ref{Jac3}). 
Due to the transformation of 
the integration variable $\beta$ to $\tau=1/\beta$ in the first term and using
$\beta$ directly as the integration variable in the second term, they
are
reduced to the analytical  
inverse-Laplace form (\ref{Laplace}) 
    for the transformation from
$\tau$ to $a$ variables
\cite{AS64}.
 Thus, one can approximately represent
the level density $\rho(E,A)$ as a superposition of the two Bessel functions of
the orders of 3/2 and 1/2,
\bea\l{rhoGEN32}
&\rho(E,A)\approx \overline{\rho}_{3/2}\left(S^{-3/2}I_{3/2}(S)-
r^{}_1 S^{-1/2}I_{1/2}(S)\right)\nonumber\\
&\mbox{with}\quad\overline{\rho}_{3/2}=a\sqrt{\frac{2\pi}{3}}~.
\eea
 Here 
\be\l{r1}
r^{}_1=\frac{\overline{\xi}U^{1/2}}{4 a^{3/2}}\approx
\frac{\pi^6K^{3/2}U^{1/2}}{3\lambda^2A^{7/6}}~\mathcal{E}_{\rm sh}~,
\ee
where $\overline{\xi}$ is given in Eq.~(\ref{xib}),
$K=A/a$, $a$ is the level density parameter, Eq.~(\ref{denpar}), and
\be\l{entrFG}
S=2\sqrt{a U}~.
\ee
This expression is associated with an entropy
in the mean field approximation because of its two clear
asymptotic limits
for large and small excitation energies, $U$ 
[both 
 asymptotic limits in terms of the level density,
$\rho(E,A)$, will be discussed 
below].
The relative contribution of the second term 
    in Eq.~(\ref{rhoGEN32})
decreases with
the shell effects, $\mathcal{E}_{\rm sh}$, inverse level density parameter,
$K$, and excitation energy, $U$.
In the case (i),  referred to below as the MMA1 approach,
up to these corrections (small $r^{}_1$), 
one arrives  approximately  at 
expression
 (11) of Ref.\ \cite{PLB}:
\be\l{rho32}
\rho(E,A) \approx \overline{\rho}_{3/2}~S^{-3/2}I_{3/2}(S)~,\qquad  \mbox{(i)}~.
\ee

\subsection{(ii) Dominating shell effects}
\l{subsec-52}

In this case, expanding the Jacobian factor $\mathcal{J}^{-1/2}$,
see Eq.~(\ref{Jac3}),  over small
$1/\xi$, 
one finds
\be\l{Jac4}
\frac{1}{\mathcal{J}^{1/2}}\approx\frac{1}{\sqrt{g(\lambda)\xi}}
\left(1-\frac{1}{2\xi}\right)~, 
\ee
where $\xi>0$, Eq.~(\ref{xiLIN}) (for $\delta E<0$).
Substituting this approximate expression for the Jacobian factor
into Eq.~(\ref{rhoE1F}) and  
 transforming the integration variable $\beta$ to $\tau=1/\beta$ in the integral
representation for the level density $\rho(E,A)$, 
we obtain by using the
inverse Laplace transformation 
(\ref{Laplace}) from  $\tau$ to $a$ 
variable:
\bea
&\rho(E,A)\approx \overline{\rho}_{5/2}
\left(S^{-5/2}I_{5/2}(S) + r^{}_2 S^{-9/2}I_{9/2}(S)\right),\l{rhoGEN52}\\
&\mbox{with}
\quad\overline{\rho}_{5/2}=
4a^2\left(\pi/6\overline{\xi}\right)^{1/2}~\l{rhobar52},
\eea
where $\overline{\xi}$ is given by Eqs.~(\ref{xib}) and (\ref{xibdE}), and
\be\l{r2}
r_{2}=\frac{2a^2}{\overline{\xi}}
\approx \frac{3 \lambda^2A^{5/3}}{2\pi^6K^2\mathcal{E}_{sh}}~.
\ee
In contrast to 
case (i), the relative contribution of
the second term 
in the r.h.s. 
of Eq.~(\ref{rhoGEN52}) [case (ii)]
has the
opposite  behavior in the values of parameters $\mathcal{E}_{\rm sh}$ and $K$,  and is
    almost independent of $U$.  Up to 
small contribution of the second term in Eq.~(\ref{rhoGEN52}),
one arrives  approximately at
\be\l{rho52}
\rho(E,A)\approx 
\overline{\rho}_{5/2}~S^{-5/2}I_{5/2}(S)~, \qquad \mbox{(ii)}~,
\ee
 where $\overline{\rho}_{5/2}$ is given by Eq.~(\ref{rhobar52}). This approximation is
    referred to below
as the MMA2 approach.

Figure \ref{fig1} shows good convergence of
      different approximations to their main term ($n=0$)
      for $\rho(E,A)$.  Here we accounted
      for the
      first ($n=1$) analytical and second ($n=2$) numerical corrections
      in the expansion
      of the Jacobian factor $\mathcal{J}^{-1/2}$ [see Eq.~(\ref{Jac2})
      for the Jacobian $\mathcal{J}$], over 
  $1/\xi$ (MMA2) and over $\xi$ (MMA1) as functions of
      the excitation energy $U$. Calculations are carried out
      for typical values of the
          parameters: 
      the inverse level density $K$, the
      relative energy shell corrections $\mathcal{E}_{\rm sh}$, and
      a large particle number $A$.
      The results of the analytical MMA1 approach,
      Eq.~(\ref{rhoGEN32}), and MMA2, Eq.~(\ref{rhoGEN52}), with 
          the first
         correction terms, are compared with those of
      Eqs.~(\ref{rho52}) and (\ref{rho32}) without
      first 
      correction terms, respectively, using 
     different values of these  parameters.
  The contributions of these corrections to the simplest analytical expressions,
  Eq.~(\ref{rho32}) and (\ref{rho52}), are 
  smaller 
  the smaller excitation energies $U$ for the MMA1 and the larger
  $U$ for the MMA2 such that
  a transition
  between the 
  approaches, Eq.~(\ref{rhoGEN32}) and (\ref{rhoGEN52}), takes
  place with increasing 
  value of $U$; see
  Fig.~\ref{fig1}. We 
   also demonstrate good convergence to
  the leading terms ($n=0$)
  by taking into account numerically the next order ($n=2$ in this figure)
      corrections in the direct calculations of the 
      integral representation (\ref{rhoE1F}).
      Such a convergence occurs for the MMA1 better for  smaller $U$
      with increasing inverse density parameter $K$
      and decreasing 
      relative energy shell correction
  $\mathcal{E}_{\rm sh}$. An opposite behavior takes place for the MMA2 approach.
  Especially, a 
  good 
  convergence with increasing 
  excitation energy $U$
  is seen 
  clearly with $n=1$ and $2$ for
  the MMA1 in panels (a) and (c); see, e.g., panel (c) for larger values of
      both $K$
  and  
  $\mathcal{E}_{\rm sh}$.

  Notice that for the case (ii) when the shell
  effects are dominating, 
  the derivatives are relatively large,
      $a^{\prime\prime}(\lambda)\lambda^2/a \gg 1$, but at the same time the shell
      corrections, $\mathcal{E}_{\rm sh}$, 
      can be small. 
     In this case, referred to below as the MMA2b approach,  we have 
     for the coefficient $\overline{\rho}_{5/2}$
  \be\l{rhobar52TF}
  \overline{\rho}_{5/2}\approx 2\sqrt{2/\pi}\lambda a^2.
\ee
Here,  in the calculation of $\overline{\rho}_{5/2}$ given 
  by Eq.~(\ref{rhobar52}),
    we used the
TF evaluation of the level density, 
$\tilde{g}\propto A/\lambda$, and its
  derivatives over $\lambda$ in the
  first equation of (\ref{xib}) for $\overline{\xi}$.

\subsection{Disappearance of shell effects with temperature}
\l{subsec-LT}

As well known (see for instance Refs.~\cite{BB03,SM76,KM79,MG17}),
with increasing temperatures
$T$, the shell component $\delta \Omega$, Eq.~(\ref{potoscparF}),
disappears exponentially as $\exp(-2\pi^2T/D_{\rm sh})$
in the potential $\Omega$ or free energy $F$, see also
Eqs.~(\ref{FESCF}) and (\ref{dFESCF}). This 
 occurs at temperatures
$T\approx D_{\rm sh}/\pi=2-3$ MeV
($D_{\rm sh}=\lambda/A^{1/3}=7-10$ MeV in heavy nuclei,
$A\approx 100-200$). For such large temperatures with excitation energies $U$, near or larger
than neutron resonances energies, 
one can approximate
the Jacobian factor $\mathcal{J}^{-1/2}$ in Eq.~(\ref{rhoE1F}) as
\be\l{Jac5}
\mathcal{J}^{-1/2}\approx \tilde{\mathcal{J}}^{-1/2}\left(1-\delta \mathcal{J}/(2\tilde{\mathcal{J}})
\right)~,
\ee
 where
$\tilde{\mathcal{J}}\approx \tilde{g}$, and
\be\l{dJac5}
\delta \mathcal{J}\approx
2 \sum^{}_{\rm PO}g^{}_{\rm PO}x^{}_{\rm PO}~\exp\left(-x^{}_{\rm PO}\right),
\ee
 and 
$x^{}_{\rm PO}=\pi t^{}_{\rm PO}/\hbar \beta$, Eq.~(\ref{dFESCF}). With this approximation,
using the transformation of 
the integration variable $\beta$ to $\tau=1/\beta$
in Eq.~(\ref{rhoE1F}), 
one can analytically take the
inverse Laplace integral 
[Eq.~(\ref{Laplace})] for the level density.  Finally, one obtains 
$\rho=\tilde{\rho}+\delta \rho$, where
\bea\l{rhoT}
&\delta \rho(E,A)=\sqrt{\frac{\pi}{2 \tilde{g}^3}}
 \sum_{\rm PO}^{} g^{}_{\rm PO}\frac{t^{}_{\rm PO}}{\hbar}
\int \frac{{\rm d} \tau}{2\pi i\tau^{3/2}}\exp\left(a_{\rm sh} \tau +
\frac{U}{\tau}\right)\nonumber\\
&=\!\!\sqrt{\frac{\pi}{2 \tilde{g}^3}}
\sum^{}_{\rm PO}g^{}_{\rm PO}\frac{t^{}_{\rm PO}}{\hbar}\left(4 a~a^{}_{\rm sh}\right)^{1/4}
S_{\rm sh}^{-1/2}I_{1/2}\left(S_{\rm sh}\right).
\eea
 Here,
$a^{}_{\rm sh}=\tilde{a}-\pi t^{}_{\rm PO}/\hbar$
is the shifted level density parameter
due to the shell effects, and $S_{\rm sh}=2\sqrt{a^{}_{\rm sh}U}$ is 
the
shifted entropy. For a major shell structure, one arrives at
\bea\l{rhoTms}
&\delta \rho(E,A)\approx
\sqrt{\frac{\pi}{2 \tilde{g}^3}}\frac{2\pi}{D_{\rm sh}}\left(4 a~a^{}_{\rm sh}\right)^{1/4}
\delta g(\lambda)
~S_{\rm sh}^{-1/2}I_{1/2}\left(S_{\rm sh}\right)\nonumber\\
&\approx
\sqrt{\frac{\pi}{2 \tilde{g}^3}}\left(\frac{2\pi}{D_{\rm sh}}\right)^3
\left(4 a~a^{}_{\rm sh}\right)^{1/4}
\delta E~
S_{\rm sh}^{-1/2}I_{1/2}\left(S_{\rm sh}\right)
\eea
[see Eq.~(\ref{dedg})], and
\be\l{ash}
a^{}_{\rm sh}\approx \tilde{a}-\frac{2\pi^2}{D_{\rm sh}}~.
\ee
Hence, the  
shifted inverse level-density parameter is  
$K=A/a~=~ \tilde{K}\left(1+\Delta K/\tilde{K} \right)$,
where the dimensionless shift is given by 
\be\l{Ksh}
\frac{\Delta K}{\tilde{K}} \approx
\frac{2\pi^2\tilde{K}}{A D_{\rm sh}}\approx\frac{2\pi^2\tilde{K}}{\lambda A^{2/3}}~.
\ee
This is approximately equal to
$\Delta K \approx 1-2$ MeV for $\tilde{K}=10$ MeV
(see Refs.~\cite{SN90,SN91,KS18,KS20})
at typical parameters $\lambda=40$ MeV and $A=100-200$
($\Delta K \approx 6-9$ MeV for $\tilde{K}=20$ MeV).
 We note that an important
shift 
in the inverse level density parameter $K$
for double magic nuclei near the neutron resonances  is due to a strong shell
effect.

  \subsection{General MMA}  
  \l{subsec-GMMA}

  All final results for the level density $\rho(E,A)$ discussed in the
  previous 
  subsections 
  of  this section
  can be approximately summarized 
  as  
  \begin{equation}\label{denbes}
    \rho \approx \rho^{}_{\tt{MMA}}(S)
    =\overline{\rho}_\nu~f_\nu(S)~,~~~f_\nu(S)=
  S^{-\nu}I_{\nu}(S)~,
  \end{equation}
  with
   corresponding
   expressions for the 
coefficient $\overline{\rho}_\nu$ (see above).
For large entropy
$S$, one finds
 \begin{equation} 
 f_\nu(S) =\frac{\exp(S)}{S^{\nu}\sqrt{2\pi S}}\left[1+\frac{1-4\nu^2}{8S}
    +\mbox{O}\left(\frac{1}{S^2}\right)\right].
 \label{rhoasgen}
\end{equation}
     At small entropy, $S \ll 1$, one obtains
     also from Eq.~(\ref{denbes})
     the finite combinatorics power
expansion \cite{St58,Er60,Ig72,Ig83}
\begin{equation}
    f_\nu(S)=
  \frac{2^{-\nu}}{\Gamma(\nu+1)}\left[1+\frac{S^2}{4(\nu+1)}+
    \mbox{O}\left(S^4\right)\right],
\label{den0gen}
\end{equation}
where $\Gamma(x)$ is the gamma function.
    This expansion over powers of 
    $S^2 \propto U$ 
    is the same 
    as that of  the 
    ``constant temperature model'' (CTM) \cite{GC65,ZK18,Ze19},
used often for the level density calculations  at
small excitation
energies $U$, but here we have it without
free parameters.

In order to clarify  Eq.~(\ref{rhoasgen}) for the MMA level density at
a large entropy, one can directly obtain a more general full SPM asymptote,
including
    the shell effects,
by taking
the integral over $\beta$ in Eq.~(\ref{rhoE1F}) 
 using
the SPM
      (see Appendix \ref{appC}).  We have
%
%%%% FIG. 2 %%%%%%%
\vspace{0.2cm}
\begin{figure*}
    \includegraphics[width=11.5cm]{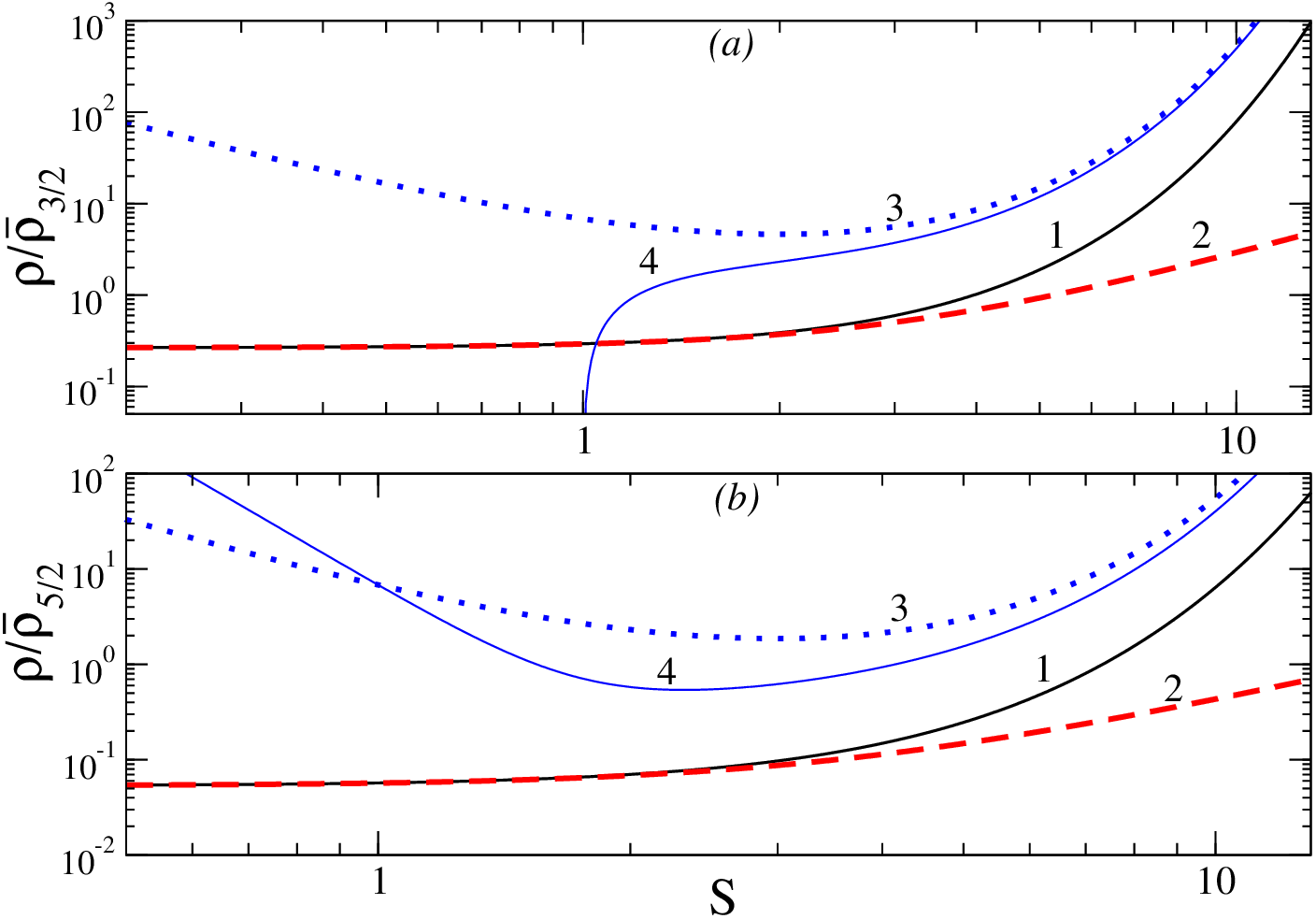}  

\vspace{-0.2cm}
\caption{{\small 
    Level density $\rho$ [Eq.~(\ref{denbes})], 
    in units of $\overline{\rho}_\nu$, with the accurate result ``1''
    (solid line),
    Eq.~(\ref{rho32}), $(a)$
    for $\nu=3/2$ [MMA1 (i)], and $(b)$, Eq.~(\ref{rho52}),
    for $\nu=5/2$ [MMA2 (ii)], 
    shown as
    functions of the entropy $S$ for different approximations:
     (1) $S \ll 1$
    (dashed lines),  Eq.~(\ref{den0gen})
    at the second order, and (2) $S \gg 1$
    (dotted and 
    thin solid lines),~ 
       Eq.~(\ref{rhoasgen}); 
       ``3'' is the main term of the expansion in powers of $1/S$, and ``4''
     is the expansion  over $1/S$ 
     up to 
     first [in $(a)$], and second [in $(b)$]
    order terms in square brackets of Eq.\ (\ref{rhoasgen}),
    respectively.}}
\label{fig2}
\end{figure*}
\be\l{SPMgen}
\rho(E,A)=
\frac{\exp(2\sqrt{aU})}{\sqrt{48}~U~\sqrt{1+\xi^\ast}}~,
\ee
where $\xi^\ast$ is $\xi$ of Eq.~(\ref{xipar})  at the saddle point
$\beta=\beta^\ast$, which is, in turn, determined by
 Eq.~(\ref{spmcondbeta}):
\be\l{par}
\xi^\ast\approx
-\frac{\pi^2T^2}{6\hbar^2}
\sum^{}_{\rm PO}t^{2}_{\rm PO}\frac{g^{}_{\rm PO}(\lambda)}{g(\lambda)} \approx 
\frac{4\pi^6 U K\mathcal{E}_{\rm sh}}{3 \lambda^2 A^{2/3}}~.
\ee
We 
took 
the factor $\mathcal{J}^{-1/2}$, 
 obtained from the Jacobian $\mathcal{J}$
of Eq.~(\ref{Jac2}), off the integral
(\ref{rhoE1F}) at
$\beta=\beta^\ast=1/T$.
The  Jacobian ratio $\xi^\ast$ of
    $\delta \mathcal{J}/\tilde{\mathcal{J}}$ at the saddle point,
$\beta=\beta^\ast$  ($\lambda=\lambda^\ast=\alpha^\ast T$ is the standard
chemical potential of the grand-canonical ensemble),
Eq.~(\ref{par}),
is the   critical quantity for these derivations.
The quantity 
$\xi^\ast$ is approximately proportional
to the semiclassical POT 
energy shell correction, $\delta E$, Eq.~(\ref{dedg}), 
through
   $\mathcal{E}_{\rm sh}$, Eq.~(\ref{xibdE}), 
the excitation energy, $U=a T^2$,
    and to a
 small 
semiclassical
    parameter $A^{-1/3}$ squared for heavy nuclei
    (see Ref.~\cite{PLB} and Appendix \ref{appA}).
  For typical values of parameters,
  $\lambda=40$ MeV, $A\approx 200$,
  and 
  $\mathcal{E}_{\rm sh}=|\delta E~A/E_{\rm \tt{ETF}}|\approx 2.0$
  \cite{BD72,MSIS12},
  one finds the approximate values of 
  $\xi^\ast\approx 0.1 - 10$ for temperatures $T \approx 0.1-1$ MeV. 
      This corresponds approximately
   to rather wide 
  excitation energies $U=0.2-20$~ MeV for $K=10$~MeV \cite{KS18}
  (and $U=0.1-10$~MeV for $K=20$~ MeV). 
  These values of $U$
      overlap 
      the interval of 
      energies of the 
      low energy states  with that of the 
      energies of states 
      significantly
  above the neutron resonances.
  In line with the SCM \cite{BD72} and ETF approaches \cite{BB03},
  these values are given 
  by 
  the realistic
    smooth energy $E_{\rm \tt{ETF}}$ for which
    the binding energy  approximately equals 
  $E_{\rm \tt{ETF}}+ \delta E$ \cite{MSIS12}.

     Accounting for the shell effects, Eq.~(\ref{SPMgen})
    is 
    more general large-excitation energy asymptote
  with respect to the well-known Bethe expression \cite{Be36}
\be\l{ldBethe}
\rho(E,A) = \frac{\exp\left(S\right)}{\sqrt{48}U}~,
\ee
 where such effects were neglected; see also
    Refs.~\cite{Er60,GC65,BM67}.
This expression can be alternatively 
obtained as the limit of Eq.~(\ref{SPMgen})
at large excitation energy,
$U \rightarrow \infty$, up to shell effects [small $\xi^\ast$
of the case (i)]. 
This asymptotic result is 
the same as 
that of 
expression (\ref{rho32}),
proportional to the Bessel function $I_\nu$ 
of the order $\nu=3/2$  [the case (i)], at the main zero-order
expansion in $1/S$; see Eq.~(\ref{rhoasgen}).
 For 
large-entropy $S$ asymptote, we 
find also that the
Bessel solution (\ref{rho52}) with $\nu=5/2$
in the case (ii) ($\xi^\ast \gg 1$)
at zero-order
expansion in $1/S$ coincides with that of the
general asymptote (\ref{SPMgen}). 
The asymptotic expressions, Eqs.~(\ref{rhoasgen}), (\ref{SPMgen}), and,
in particular,
(\ref{ldBethe}),  for the level density
are obviously divergent at $U\rightarrow 0$, in
contrast to the finite MMA limit (\ref{den0gen}) for the level density;
see Eq.~(\ref{denbes}) and, e.g., Eqs.~(\ref{rho32}) and (\ref{rho52}).

Our MMA results will be compared also with the popular Fermi gas 
(FG) approximation to the level density  $\rho(E,N,Z)$ as a
function of the neutron $N$ and proton $Z$ numbers near the $\beta $
stability line, $(N-Z)^2/A^2\ll 1$
\cite{Er60,GC65,EB09}:
\begin{equation}
  \rho(E,N,Z)=
\frac{\sqrt{\pi}}{12 a^{1/4}
   U^{5/4}}~
  \exp\left(2\sqrt{a U}\right)~. 
\label{intFGSS}
\end{equation}

 Notice that in all our calculations of the statistical level density,
$\rho(E,A)$ [also $\rho(E,N,Z)$, Eq.~(\ref{intFGSS})],
we did not use a popular assumption of small spins at large
excitation energy which is valid for the neutron resonances.
For typical values of spin $I \simg 10$, 
  moment of inertia
$\Theta \approx \Theta_{\rm \tt{TF}}
 \approx 2\mu R^2 A/5$,
Eq.~(\ref{rigMIpar}),  radius
$R=r^{}_0A^{1/3}$,  with $r_0=1.14 $ fm, and particle number $A \siml 200$,
one finds that,  for large entropy, the applicability condition
(\ref{condI2}) 
is not
strictly speaking valid. 
In these estimates, the corresponding excitation energies $U$ of LESs are
essentially smaller
than neutron resonance energies.
    However, near neutron resonances the excitation energies $U$
    are large,
        spins  are small, and Eq.~(\ref{intFGSS})
    is well justified. 

  We should also emphasize that the MMA1
approximation for the level density,
  $\rho(E,A)$,
  Eq.~(\ref{rho32}), and the Fermi gas approximation, Eq.~(\ref{ldBethe})
  can be also applied for
  large excitation energies $U$, with respect to the collective
  rotational  excitations,
  if one can neglect shell effects, $\xi^{\ast}\ll 1$.      
Thus, with increasing temperature $T \simg 1$ MeV (if it exists),
or excitation energy $U$, 
where the shell effects are yet significant,
    one first obtains 
    the asymptotical expression (\ref{SPMgen}) at $\xi^\ast\gg 1$, i.e.,
    the asymptote of
    Eq.~(\ref{rho52}). Then, with further increasing temperature to about 2-3 MeV
     with the disappearance of shell effects (section \ref{subsec-LT}),
     one 
    gets the transition to
    the Bethe formula,  i.e., the  large entropy asymptote (\ref{ldBethe})
    of Eq.~(\ref{rho32}).

      %%%% FIG. 3 %%%%%%%
\vspace{0.2cm}
\begin{figure*}
       \includegraphics[width=11.5cm]{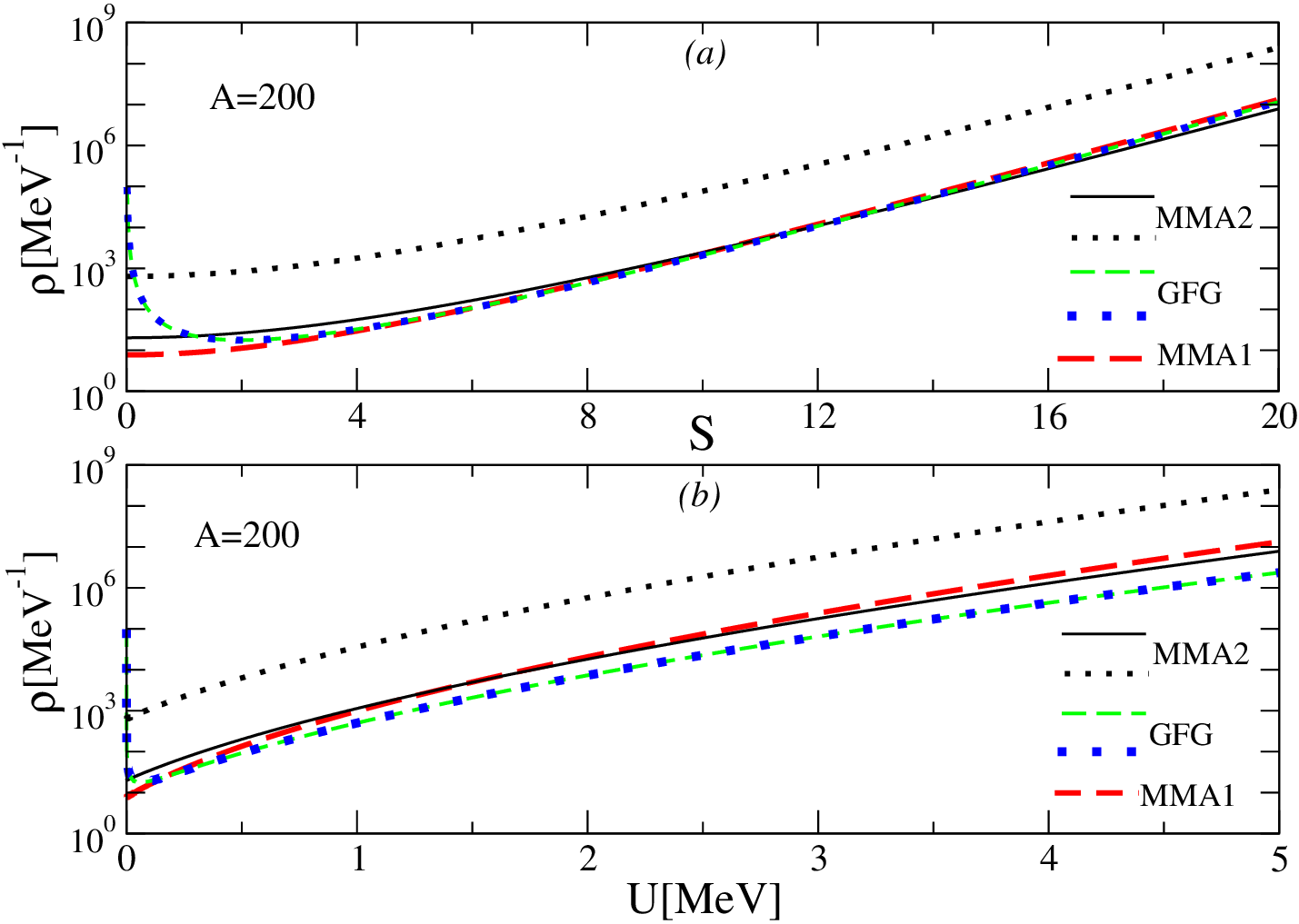} 

\vspace{-0.2cm}
\caption{{\small 
    MMA level density $\rho$ [Eq.~(\ref{denbes})] in units of MeV$^{-1}$
    as function of the entropy $S$ (a), and excitation energy $U$,
    in units of MeV (b).  The black solid and dotted lines  are the MMA2 
        approach
    for 
        $\mathcal{E}_{\rm sh}=2.0$ and $0.002$,
    Eq.~(\ref{rho52}), respectively. 
   Green dashed and blue dotted lines
    are the general Fermi gas (GFG) approach,
    Eq.~(\ref{SPMgen}), for the  
    same
    values of $\mathcal{E}_{\rm sh}$,
    respectively. The red dashed line
    is the MMA1, Eq.~(\ref{rho32});  in (b) $K=10$ MeV,
    of the order of the ETF value of
    Ref.~\cite{KS18}.
}}
\label{fig3}
\end{figure*}

In Fig.~\ref{fig2} 
we show
the level density dependence
$\rho(S)$, Eq.~(\ref{denbes}), 
for $\nu=3/2$ in $(a)$ and $\nu=5/2$ in $(b)$, 
on the
entropy variable $S$ with
the corresponding asymptote. In this figure, 
small [$S\ll 1$, Eq.~(\ref{den0gen})] and
large [$S\gg 1$, Eq.~(\ref{rhoasgen})] entropy $S$ behaviors 
are presented. For small $S\ll 1$ expansion we take into account the quadratic
approximation ``2'', where $S^2 \propto U$, that is  the same as in the linear
expansion within the
CTM \cite{GC65,ZK18}. For large $S\gg 1 $
we neglected the corrections of the inverse power entropy expansion of the
preexponent factor
in square brackets of Eq.\  (\ref{rhoasgen}), lines ``3'', and took into account
the 
corrections of the
first [$\nu=3/2$, $(a)$] and up to second [$\nu=5/2$, $(b)$]
    order in $1/S$ (thin solid lines ``4'')
to show  
their slow convergence
to the accurate 
MMA result ``1'' (\ref{denbes}).
It is interesting to find almost a 
constant shift of the results of the simplest, $\rho \propto \exp(S)/S^{\nu+1/2}$,
SPM 
asymptotic approximation at
large $S$ (dotted lines ``3'')
with respect to 
the accurate  
MMA results of Eq.~(\ref{denbes}) (solid lines ``1'').
This 
may clarify
one of the phenomenological
models, e.g., the 
back-shifted
Fermi-gas 
(BSFG) model for the level density \cite{DS73,So90,EB09}.

Figure \ref{fig3} shows the shell effects in the main approximations
derived in this section, Eqs.(\ref{rho32}), (\ref{rho52}), and (\ref{SPMgen}),
taking two essentially different values of finite
$\mathcal{E}_{\rm sh}=2.0$
and much smaller
$0.002$,
between which  
one can find
basically those given by 
Ref.~\cite{MSIS12}. For convenience, we show these results as functions
of the entropy $S$ in panel (a), and the excitation energy $U$
 in  panel (b),
taking the value of
the averaged inverse density parameter $K$  found in
Ref.~\cite{KS18}; see also Ref.~\cite{KS20}. As expected,
the shell effect is very strong for the MMA2 approach,
as can be seen from the 
difference between solid and dotted black
lines\footnote{The dotted black line is very close
      to the explicit analytical limit (\ref{rhobar52TF})
      of $\overline{\rho}_{5/2}$, Eq.~(\ref{rhobar52}),
      for the MMA2 equation (\ref{rho52}),
      see also Eq.~(\ref{rhobar52TF}).} 
depending on the
second derivatives of strong oscillating functions of $\lambda$,
$a''(\lambda)\approx \delta a''\propto \delta g''(\lambda) $
[see Appendix \ref{appA} around Eq.~(\ref{d2g}) and Sec.~\ref{sec-MMAas}
below
Eq.~(\ref{dedg})].
 This
is not the case for the full SPM  asymptotic GFG, Eq.~(\ref{SPMgen}), for which
this effect is very small.
As seen from this figure, the MMA1, Eq.~(\ref{rho32}), independently of $\mathcal{E}_{\rm sh}$,
 converges rapidly 
to the 
GFG with increasing
  excitation energy $U$ 
as well as to the Bethe formula
(\ref{ldBethe}). 
They all coincide at small values of
$U$, about 0.5 MeV, 
 particularly for $\mathcal{E}_{\rm sh}=0.002$.
The Bethe approach is 
 very close everywhere to the GFG 
line at $\mathcal{E}_{\rm sh}=0.002$
and therefore,
 it is not shown in this figure. Notice also that MMA2
at this small $\mathcal{E}_{\rm sh}$ is also close to the MMA1 
    everywhere. 
Again, one can see that the MMA1 and MMA2 have no divergence at zero excitation energy
limit, $U \rightarrow 0$, 
 while the full SPM asymptotic GFG, Eq.~(\ref{SPMgen}),  and, in particular, 
 the Bethe approach, Eq.~(\ref{ldBethe}), both diverge at $U \rightarrow 0$.
 
%\end{document}
  \subsection{The spin-dependent level density}
  \l{subsec-I}

  Assuming that there are no external forces acting on an
  axially symmetrical 
   nuclear system,
  the total angular momentum $I$ and its projection $M$ on a
  space-fixed axis 
  are conserved, and states with a given energy $E$
  and spin $I$ are $2I+1$ degenerated. As shown in Appendix \ref{appB}, 
      for the ``parallel''
      rotation around the symmetry axis $Oz$, i.e., an alignment of the
      individual  angular momenta of the particle  along $Oz$ (see Ref.~\cite{KM79}
      for the spherical case), in
      contrast to the ``perpendicular-to-axis $Oz$'' collective
      rotation 
      (see, e.g., Ref.~\cite{GM21}), 
  one can derive the level density $\rho(E,A,M) $ within the MMA
  approach in the same analytical form
  as for the $\rho(E,A)$,  Eq.~(\ref{denbes}):
  \be\l{denbes1}
  \rho^{}_{\rm \tt{MMA}}(E,A,M)\approx \overline{\rho}_{\nu}f_\nu(S)~,\qquad
  \mbox{with}\qquad\nu=2,3~,
  \ee
  where 
  \be\l{conM2}
  \overline{\rho}_2=
  \hbar~
  \left(\frac{2a^{3}}{3\Theta}\right)^{1/2},\quad \nu=2~~~\mbox(i)~,
    \ee
   and
  \be\l{conM3}
\overline{\rho}_3
=
\hbar \lambda~
\left(\frac{8a^5}{\pi^2\Theta}\right)^{1/2},\quad \nu=3~~~(ii)~.
  \ee
  In Eq.~(\ref{denbes1}), the argument of the Bessel-like 
      function, $f_{\nu}(S)\propto I_\nu(S)$,
  Eq.~(\ref{denbes}), is the
  entropy $S(E,A,M)$,
  Eq.~(\ref{entrFG}), with  the $M$-dependent excitation energy $U$.
    Indeed, in the 
  adiabatic mean-field approximation, 
  the level density parameter $a$ in Eq.~(\ref{entrFG}) 
is given by Eq.~(\ref{daF}). 
For the intrinsic excitation energy $U$ in Eq.~(\ref{entrFG}), 
one finds
  \be\l{Eex}
  U=E-E_{0}-\frac12 \Theta~\omega^2~,~~~~\omega=\frac{\hbar M}{\Theta}~,
  \ee
  where, $E_{0}=\tilde{E} +\delta E$,
  is the same intrinsic (nonrotating) shell-structure energy as in
  Eq.~(\ref{OmadF}). 
 With the help of the conservation equation (\ref{conseq})
for the  saddle point,
$\kappa^\ast=\hbar \omega \beta$, we eliminated
the rotation frequency $\omega$, 
 obtaining the second equation in Eq.~(\ref{Eex}); see Appendix \ref{appB}.
  For the
 moment of inertia (MI)
  $\Theta$
  one has a similar SCM decomposition:
  \be\l{MI}
  \Theta=\tilde{\Theta} + \delta \Theta~,
  \ee
  where $\tilde{\Theta}$ is the (E)TF MI
      component which can be approximated largely by the
      TF expression, Eq.~(\ref{rigMIpar}), and $\delta \Theta$ is the MI
      shell correction which can be presented finally for the spherically symmetric mean field
      by Eq.~(\ref{MIpar}).
    As mentioned above, Eqs.~(\ref{denbes1})-(\ref{MI}) are  valid for the
   ``parallel'' rotation (an alignment
  of nucleons' angular momenta along the symmetry axis $Oz$); 
  see
  Appendix \ref{appB} for the specific derivations by assuming
 a spherical symmetry of the potential.
   In these derivations we used Eq.~(\ref{Eex}) for the excitation  energy $U$,
   Eq.~(\ref{parfun}) for the partition function and Eqs.~(\ref{GCEpot}) and (\ref{SCMpot})
  for the potential
$\Omega(\beta,\lambda,\omega)$.
   In the  evaluations of the Jacobian,  $\mathcal{J}$, one can neglect
      shell corrections, in contrast to the  evaluations of  
      the entropy $S$ in the function $f_\nu(S)$. In the derivations of
      Eqs.~(\ref{conM2}) for $\overline{\rho}^{}_{2}$ and (\ref{conM3}) for
      $\overline{\rho}^{}_{3}$, we obtained the Jacobian components,
   $\tilde{\mathcal{J}}$ for the case (i) and $\delta \mathcal{J}$
      for the case (ii), both under the assumption of an axially symmetric mean field
       (see Appendix \ref{appB}).
  For the
  Jacobian calculations, one can finally use
   the (E)TF approximation in the case (i),
   $\Theta \approx \tilde{\Theta}$.
  The Jacobian $\mathcal{J}$ in the case (ii) 
  can be approximated by Eq.~ (\ref{Jacsph}).  As a 
  result, one may accurately
use the (E)TF approximation $\Theta \approx \tilde{\Theta}$ in Eqs.~(\ref{conM2}) and (\ref{conM3})
  for the coefficients $\overline{\rho}^{}_{2}$ and $\overline{\rho}^{}_{3}$.

Note that there is no divergence of the level density
$\rho(E,A,M)$ [Eq.~(\ref{denbes1})] in the limit $U\rightarrow 0$,
 Eq.~(\ref{den0gen}), in contrast to
    the standard results of the full SPM within the Fermi gas model.
    The latter
is  associated with the leading term in
expansion
(\ref{rhoasgen}) of the Bessel-like function $f_\nu(S)$.

    Equation (\ref{denbes1}),  with $M=\mathcal{K}$,  if it exists,
    can be used for the calculations of
    the level density 
$\rho(E,A,\mathcal{K})$, where $\mathcal{K}$ is the specific projection of the total
    angular momentum ${\bf I}$ on the symmetry axis of the axially symmetric potential \cite{MK79}
    (K in notations of Ref.~\cite{BM75}).
 We note that it is common to use in application 
\cite{Be36,Er60,BM67} the level density
    dependence on the spin $I$,
$\rho(E,A,I)$.
We will consider here only the academic axially symmetric potential
case which can be realized practically for
the spherical  or axial symmetry of a mean nuclear field  for the
    ``parallel'' rotation mentioned above.
Using Eq.\ (\ref{denbes1}), 
under the same assumption of a closed rotating system 
and, therefore,  with
conservation of the integrals of motion,  the spin $I$
and its projection $M$ on the space-fixed axis,
one can calculate the corresponding
 spin-dependent level density $\rho(E,A,I)$ 
  for a 
  given energy $E$, particle number $A$, and total
angular momentum $I$ 
 by employing  the Bethe formula \cite{Be36,BM67,Ig83,So90},
\bea \l{denEAIgen}
\rho(E,A,I)&=&\rho(E,A,M=I) - \rho(E,A,M=I+1)\nonumber\\
&\approx&
-\left(\frac{\partial \rho(E,A,M)}{\partial M}\right)^{}_{M=I+1/2}~.
\eea
For this
level density, $\rho(E,A,I)$, 
one 
obtains from Eqs.~(\ref{denbes1}) and (\ref{Eex}),
\be 
\rho^{}_{\rm \tt{MMA}}(E,A,I) \approx
\frac{a\overline{\rho}_{\nu}\hbar^2(2I+1)}{\Theta}f_{\nu+1}(S)
\l{denbesI}~,
\ee
where $S$ is given by Eq.~(\ref{entrFG})
with the excitation energy (\ref{Eex}), and $\nu$ equals 2 and 3,
in correspondence with Eq.~(\ref{denbes1}).
The multiplier $2I+1$ in
Eq.~(\ref{denbesI}) appears because of the substitution $M=I+1/2$ into
the derivative in
Eq.~(\ref{denEAIgen}). In order to obtain
the approximate MMA total level density $\rho(E,A)$
from the spin-dependent level density $\rho(E,A,I)$ we can multiply
Eq.~(\ref{denbesI}) by the spin
 degeneracy factor $2 I+1$ and integrate (sum)
over all spins $I$.

Using the expansion of the Bessel functions in Eq.~(\ref{denbesI})
over the argument
$S$
for $S\ll 1$ [Eq.~(\ref{den0gen})] 
one finds the finite  combinatorics
 expression. 
For large $S$ [large excitation energy, $aU\gg 1$,
Eq.~(\ref{rhoasgen})], one obtains from
Eq.~(\ref{denbesI}) the asymptotic Fermi gas expansion.
Again, the main term in the expansion for large $S$,
Eq.~(\ref{rhoasgen}), coincides with the full SPM limit to the
inverse Laplace integrations in Eq.\ (\ref{dengen}).
For small angular momentum $I$ and large excitation energy $U_0=E-E_{0}$,
so that
\be\l{Iexp}
\frac{E_{\rm rot}}{U_0}\approx\frac{I(I+1)\hbar^2}{2\Theta~U_0} \ll 1~,
\ee
one finds the standard separation of the level density,
 $\rho^{}_{\rm \tt{MMA}}(E,A,I)$, into the
product of the dimensionless
spin-dependent Gaussian multiplier, $\mathcal{R}(I)$, and 
another spin-independent factor. Finally, for the case (i) ($\nu=2$),
one finds 
\begin{equation}
  \rho^{}_{\rm \tt{MMA}}(E,A,I) \approx
 \frac{\overline{\rho}^{}_{2}~ \mathcal{R}(I)~\exp\left(2\sqrt{a U_0}\right)}{
16\sqrt{\pi}~(a U_0)^{5/4}} \quad (i)~.
\label{rhoIexp2}
\end{equation} 
The spin-dependent factor $\mathcal{R}(I)$ is given by
\begin{equation}
\mathcal{R}(I)=\frac{2I+1}{q^2}
\exp\left(-\frac{I(I+1)}{2q^2}\right)~,
\label{qIexp}
\end{equation} 
where $q^2=\Theta\sqrt{U_0/a}/\hbar^2$ is the dimensionless spin
dispersion. This dispersion $q$ at the saddle point,
$\beta^\ast=1/T=\sqrt{a/U_0}$, is the standard spin dispersion
$\Theta T/\hbar^2$; see Refs.~\cite{Be36,Er60}.
Similarly, for the $\nu=3$ (ii) case one obtains
\begin{equation}
  \rho^{}_{\rm \tt{MMA}}(E,A,I) \approx
   \frac{\overline{\rho}^{}_{3}~ \mathcal{R}(I)~\exp\left(2\sqrt{aU_0}\right)}{
32\sqrt{\pi}~(a U_0)^{7/4}}
 \quad (ii)~.
\label{rhoIexp3}
\end{equation} 
Note that the power dependence of the preexponent factor of the
   level density $\rho(E,A,I) $ on  the
   excitation  energy, $U_0=E-E_0$, differs from that of $\rho(E,A,M)$;
   see 
   Eqs.~(\ref{denbes1}) and (\ref{rhoasgen}). The exponential
   dependence, $\rho \propto \exp(2\sqrt{a(E-E_0)})$, for large excitation
   energy $E-E_0$ is the same for $\nu=2 $ (i) and $3$ (ii),
   but the pre-exponent factor is different; cf. Eqs.~(\ref{rhoIexp2})
   and (\ref{rhoIexp3}).
    A small angular momentum $I$ means that the 
       condition of Eq.~(\ref{Iexp})  
  was applied.  Equations (\ref{rhoIexp2}) and (\ref{rhoIexp3}) with
   Eq.~(\ref{qIexp}),  are valid 
   for excited states 
    within approximately the condition 
   $1/\tilde{g} \ll U \ll \lambda$;  see
    Eq.~(\ref{condU}).
     For relatively small
    spins [Eq.\ (\ref{Iexp})]  we have 
    the so-called 
    small-spins Fermi-gas
model (see, e.g., Refs.\ \cite{Be36,Er60,GC65,BM67,Ig83,So90,KS20}).

General derivations of equations applicable for axially symmetric systems
(a ``parallel'' rotation) in
this section are specified in Appendix \ref{appB} by
using the spherical potential
to present  explicitly  the expressions for the 
shell correction components of several POT quantities.
However, the results for the
spin-dependent level
density, $~\rho(E,A,I)$
in this section, Eqs. (\ref{denbesI})-(\ref{rhoIexp3}), cannot be immediately
applied for comparison with the available experimental data on
rotational bands  in the collective rotation of a deformed nucleus.
They are presented within the unified rotation model \cite{BM75}
in terms of the spin $I$ and its projection $\mathcal{K}$ to the internal symmetry
axis for the deformed
nuclei. 
We are going to use the ideas of
Refs.~\cite{Bj74,BM75,Gr13,Gr19,Ju98} (see also Refs.~\cite{Ig83,So90}) concerning another
definition of the
spin-dependent level density $\rho(E,A,I)$ in terms of the
intrinsic level
density and collective rotation (and vibration) enhancement in 
a forthcoming work.  The level density
$\rho(E,A,\mathcal{K})$, e.g., Eq.~(\ref{denbes1}) at $M=\mathcal{K} $, depending on  the spin
projection $\mathcal{K}$ on the symmetry axis of an axially-symmetric deformed nucleus,
can be helpful in this work. 

%
%%%% FIG. 4  %%%%%%%
\vspace{-0.4cm}
\begin{figure*}
  \vskip1mm
 \includegraphics[width=12.0cm]{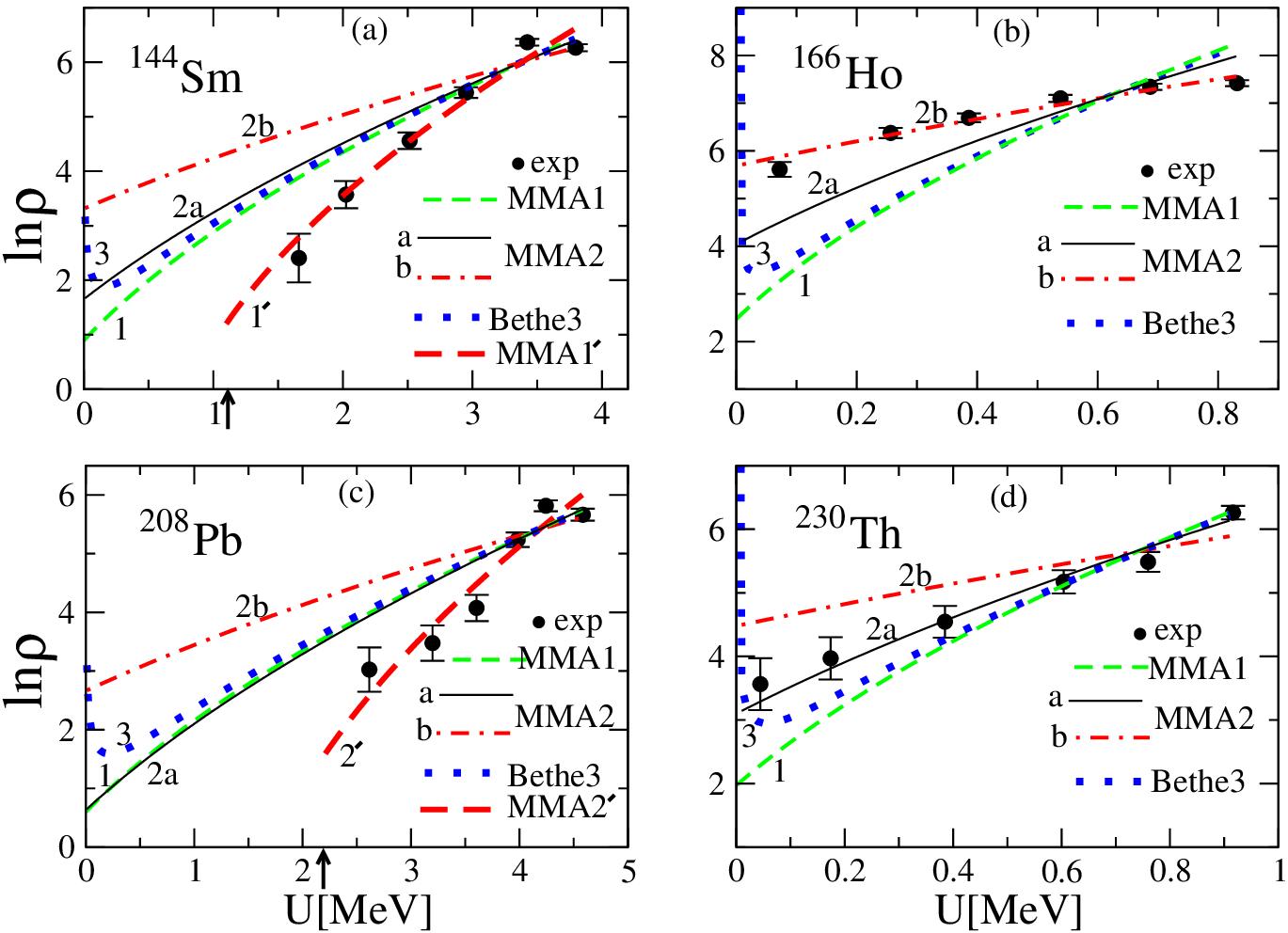} 

 \vskip-3mm\caption{{\small 
    Level density, $\mbox{ln}\rho(E,A)$, is obtained for
    low energy states in 
    $^{144}$Sm (a),
    $^{166}$Ho (b), $^{208}$Pb (c) and
    $^{230}$Th (d)
    within different
    approximations: The MMA 
    dashed green line ``1'',
    Eq.~(\ref{rho32}); the MMA solid black line ``2a'', Eq.~(\ref{rho52}),
    at the relative realistic  shell correction
    $\mathcal{E}_{\rm sh}$ \cite{MSIS12}; the MMA dash-dotted red ``2b'',
    Eq.~(\ref{rho52}) at an extremely
    small $\mathcal{E}_{\rm sh}$, 
    Eq.~(\ref{rho52}) with (\ref{rhobar52TF}); 
    and
       the Fermi gas Bethe3 
 rare blue dotted line, 
      Eq.~(\ref{ldBethe}).
      The realistic values of $
     \mathcal{E}_{\rm sh}$= 0.37 (a), 0.50 (b),
     1.77 (c), and 0.55 (d) for MMA2 are    
     taken from Ref.~\cite{MSIS12}
     (the chemical potential
      $\lambda=40$ MeV, independent of particle numbers).
     Heavy
     dashed red lines test shifts of the
     excitation energies $U$ for MMA1 and MMA2a
     by +$1.1$ and $+2.2$ MeV 
     in  $^{144}$Sm and $^{208}$Pb, respectively,  which  are
         due, presumably, to the pairing
     condensation energy shown by arrows
     in the panels $(a)$ and  $(c)$,
         as explained in the text and Table \ref{table-1}.
     Experimental dots (with error bars,
     $\Delta \rho_i/\rho_i=1/\sqrt{N_i}$)
     are obtained directly from the 
          excitation 
          states (with spins and their degeneracies)
    spectrum \cite{ENSDFdatabase} in shown nuclei 
     (Table \ref{table-1})
    by using the sample method where the sample lengths
    $U_s=0.45 (a), 0.15 (b) ,
    0.34 (c)$, and $0.17 (d)$~MeV are found on
    the plateau condition over the inverse
    level density parameter
    $K$.
}}
\label{fig4}
\end{figure*}

\section{Discussion of the results}
\l{sec-disc}

 In Fig.\ \ref{fig4} 
and Table~\ref{table-1} we present results of theoretical calculations
of   
the statistical level density $\rho(E,A)$ (in logarithms) within the MMA, 
    Eq.~(\ref{denbes}),
  and Bethe, Eq.~(\ref{ldBethe}), approaches
as functions of the
excitation energy $U$ and compared  
with experimental
data. The results of the popular
     FG approach, Eq.~(\ref{intFGSS}), and our GFG,
        Eq.~(\ref{SPMgen}), 
 are  very close to those of the Bethe
 approximation and, therefore, they
are presented only in Table \ref{table-1}.
All of the  presented results 
are calculated  by using 
the values of the inverse level density parameter $K$  obtained
from their least mean-square 
fits (LMSF)
to experimental data for several nuclei.
 The 
data shown by dots with error bars in Fig.~\ref{fig4}
are obtained for the statistical level density $\rho(E,A)$
from
the experimental data for the excitation energies $U$ and spins $I$ of the states spectra
\cite{ENSDFdatabase}
by using the sample method:
    $\rho_i^{\rm exp}=N_i/U_s$, where $N_i$
is the number of states in the $i$th sample, $i=1,2,...,N_{\rm tot}$;
see,
e.g., Refs.\ \cite{So90,LLv5}. 
    The dots are plotted at mean positions
$U_i$ of the excitation energies for each $i$th sample.
Convergence of the sample method over the equivalent sample-length
parameter $U_s$ of the statistical averaging 
was studied 
    under 
    statistical plateau conditions, 
for all plots
in Fig.~\ref{fig4}.
The sample lengths $U_s$  play  
a role which is similar to that of 
averaging parameters in the 
Strutinsky smoothing procedure 
for the SCM calculations of the averaged s.p. level density
\cite{St67,BD72}.
This plateau means almost constant 
value of the 
physical parameter $K$ within 
 large enough
    energy intervals
    $U_s$.
A sufficiently good plateau was obtained in a wide range around the
values near $U_s$ for nuclei presented in 
Fig.~\ref{fig4} and Table~\ref{table-1} 
\cite{ENSDFdatabase,HJ16}. Some values of $U_s$ are given in the caption 
    of Fig.~\ref{fig4}.
Therefore, the results
of Table
\ref{table-1}, 
calculated at the same  values of the found plateau, 
 do not depend, 
with the statistical accuracy, on the averaging
parameter $U_s$ 
within the plateau.  This is  similar to the results 
that the energy and density
shell corrections are independent of the smoothing parameters
in the SCM.
 The {\it statistical condition, $N_i\gg 1$ at
  $N_{\rm tot}\gg 1$,}
 determines the accuracy
    of our calculations. 
Microscopic details are neglected under these conditions, but one obtains more
simple, general, and analytical results,
in contrast to a micro-canonical approach. 
    As in the SCM, in our calculations
    by the sample method  
with  good plateau values for the sample lengths $U_s$ (see
        the caption of the
        Fig.~\ref{fig4}),  
            one obtains a
            sufficiently smooth 
statistical level density as a function of the excitation energy $U$. 
    We 
 require such a smooth function  because 
the statistical fluctuations are neglected 
        in our theoretical derivations.   
        %
%%%%% Table 1 *****************
 \begin{table*}[pt]
   \begin{center}
 \begin{tabular}{|c|c|c|c|c|c|c|c|c|c|}
 \hline
 Nuclei & $\langle\Delta \rho_i/\rho_i \rangle$
 &$\mathcal{E}_{\rm {sh}}$& %$\langle\Delta \rho_i/\rho_i \rangle$ &
 Version %($\mathcal{E}_{\rm sh}$)
 & K~[MeV]
 & $\sigma$ && Version %($\mathcal{E}_{\rm sh}$)
 & K~[MeV]
 & $\sigma$  \\
 %\hline
 \hline
 Sm-144&~0.18~&~0.37
 & MMA2b
 & 40.3
 &~5.1~&
 %\hline
 & MMA1$^{\ast}$ 
 & 22.7 (16.7$^\ast$)
 &~3.6 (3.3$^\ast$)~\\
  & & & GFG 
 & 21.8
 & 3.8& 
 & MMA2a 
 &22.1 
 &~3.9
 \\
& & &Bethe 
& 23.2
&~3.7~&
& FG 
& 19.7
&~3.6~\\
\hline
Sm-148&~0.17~&~0.12~
& MMA2b
& 32.5
&~5.2~&
& MMA1
& 16.8
&~1.5\\
& & & GFG 
& 16.9
&~1.7~&
& MMA2a 
& 19.3
&~3.0\\
& & &Bethe  
& 17.2
&~1.7~&
& FG  
& 14.6
&~1.6\\
\hline
Ho-166&~0.09~ &~0.50 
& MMA2b 
& 17.5
&~1.6~&
& MMA1
& 5.4
&~12.3~\\
& & & GFG 
& 5.5
&~11.1~&
&  MMA2a 
& 7.1
&~7.0\\
& & &Bethe 
& 5.6
&~11.2~&
&  FG
& 4.7
&~11.5\\
\hline
Pb-208~&~0.20~&~1.77~
& MMA2b 
& 70.1
&~3.8~&
& MMA1 
& 43.9
&~3.1\\
& & &GFG 
& 36.5
& 3.1&
& MMA2a$^\ast$ 
&34.9 (21.9$^\ast$)
&~3.0 (2.4$^\ast$)\\
& & &Bethe 
& 45.1
&~3.2~&
&  FG 
& 38.2
&~3.1\\
\hline
Th-230&~0.24~&~0.55~ 
& MMA2b 
& 36.8
&~2.6~&
& MMA1 
& 12.3
&~2.1\\
& & &GFG 
& 12.7
&~1.3&
& MMA2a 
& 14.9
&~0.9\\
& & & Bethe  
& 12.9
&~1.3~&
& FG
& 10.8 
&~
1.3\\
\hline
\end{tabular}

\vspace{-0.2cm}
\caption{{\small
     The maximal mean 
    errors (second column)
    in the statistical distribution of the states over the
    samples,
        $\langle\Delta \rho_i/\rho_i\rangle=\langle 1/\sqrt{N_i}\rangle $,
    in nuclei (first column)
    from Ref.~\cite{ENSDFdatabase}; the relative energy shell corrections
    $\mathcal{E}_{\rm sh}$,
    Eq.~(\ref{xibdE}})
    (third column, from Ref.~\cite{MSIS12});
    the
    inverse level density parameter $K$ (fifth and eighth columns),
    found by the LMSF with the
    precision of the standard expression for $\sigma$, Eq.~(\ref{chi}),
    (sixth and ninth columns)  by using the sample method and experimental data
    from Ref.~\cite{ENSDFdatabase},
    are shown
    for the version of the approximation in the
        fourth and seventh columns
        at the relative shell corrections $\mathcal{E}_{\rm sh}$  of
        the third column.
     The MMA1 and MMA2b (with the same notations for different MMA as in
    Fig.~\ref{fig4}) are
    MMA approaches
    (\ref{rho32}) ($\nu=3/2$) and (\ref{rho52}) ($\nu=5/2$
    at extremely small $\mathcal{E}_{\rm sh}$); GFG is the
    general full Fermi gas  asymptote (\ref{SPMgen}). The MMA2a
    is a more general MMA,
    Eq.~(\ref{rho52}), at different relative shell corrections
    $\mathcal{E}_{\rm sh}$ \cite{MSIS12}. 
    The asterisks denote 
     the 
    MMA1 and MMA2a approaches which are shifted
    along the excitation energy
    $U$
    axis
    by the assumed pairing condensation energy
        $E_{\rm cond}\approx 1.1$ and $2.2$ MeV, $U \rightarrow U-E_{\rm cond}$,
    for $^{144}$Sm and $^{208}$Pb 
        as shown in parentheses, respectively
    (see Sec.~\ref{sec-disc}).
    Bethe
    [Eq.~(\ref{ldBethe})] and FG [Eq.~(\ref{intFGSS})]  approaches are the
    same as in Refs.~\cite{Be36,Er60,GC65}. 
}
\label{table-1}
\end{center}
\end{table*}

The relative quantity $\sigma$ of the standard LMSF
(see 
Table \ref{table-1}),  which
determines the applicability 
of
the theoretical 
approximations, $\rho(U_i)$ (Sec.~\ref{sec-MMAas}) 
for the description of the experimental
data \cite{ENSDFdatabase} $\rho_i^{\rm exp}$,  is given by
\be\l{chi}
\sigma^2=\frac{\chi^2}{N_{\rm tot}-1}~,\quad
\chi^2=\sum_{i=1}^{N_{\rm tot}}
\frac{(y(U_i)-y^{\rm exp}_i)^2}{(\Delta y_i)^2}~,
\ee
 where $y=\ln\rho$ and $\Delta y_i \approx
    1/\sqrt{N_i}$. For the theoretical
        approaches one has the conditions of the applicability 
       assumed in
their derivations.
We consider the commonly accepted Fermi gas  asymptote
\cite{Be36,Er60,BM67,LLv5,Ig83,So90} 
for large
excitation energies $U$;
    see 
    the Bethe [Eq.~(\ref{ldBethe})] and 
     FG [Eq.~(\ref{intFGSS})] approaches,
    cf.\ with Eq.~(\ref{rhoasgen})
    and 
    our GFG (with shell effects) expression (\ref{SPMgen}).
     In a forthcoming work we will
    use the 
   asymptote of Eqs. (\ref{rhoasgen}) and (\ref{SPMgen}),
        and the sample method for
 evaluations of the statistical accuracy of the experimental data
at relatively large excitation energies (near and higher than neutron
 resonances).
It is especially helpful in the case of low-resolution
 dense states 
 at sufficiently large excitation energies.
  The examination using the value of  
  $\sigma$ 
     obtained 
    by the LMSF is 
    an additional procedure for examining 
    these theoretical conditions,
   using 
    the available experimental data.  
Notice also that application of the sample method  
 in 
determining
the experimental statistically
averaged level density
from the
nuclear spectra in terms of 
$\sigma^2$
differs essentially from the methods
 employed
in previous works (see, e.g., Ref.\ \cite{EB09})
by using the
statistical averaging of the nuclear level density and
accounting for the spin degeneracies of the 
excited states.
We do not use 
empiric free parameters in all of our calculations, in particular, for
the FG results shown in 
    Table~\ref{table-1}.   
        The
commonly accepted nonlinear
FG asymptote (\ref{rhoasgen}) could be 
 a 
    critical (necessary but, of course, not sufficient)
  theoretical guide which, with a given statistical accuracy,
        is
    helpful for understanding
spectrum completeness 
of the experimental data at large excitation energies where the spectrum
is very dense.

Figure \ref{fig4} 
shows the two opposite situations
concerning the state distributions as functions of the excitation energy $U$.
We 
show results for the spherical magic $^{144}$Sm (a) and double magic
    $^{208}$Pb (c) nuclei with 
maximal (in absolute value but negative) shell 
correction energies, in terms of the positive,
$~\mathcal{E}_{\rm sh}$; see  Table \ref{table-1} and Ref.~\cite{MSIS12}. 
In these nuclei, there are 
almost no states with extremely low
excitation energies in 
the range of $U \siml 1-2$ MeV
\cite{ENSDFdatabase}. 
In Table \ref{table-1}, we present also results for the deformed
 nucleus $^{148}$Sm where only a few
levels 
exist 
in such a
range which yields
entropies $S \siml 1$.
For the significantly deformed 
nucleus $^{166}$Ho,
with intermediate values of $\mathcal{E}_{\rm sh}$ between minimum and maximum
[Fig.~\ref{fig4}(b)],
one finds the opposite situation when
there 
are many such LESs. An intermediate number of
 LESs is observed, e.g.,
in another deformed nucleus, $^{230}$Th [Fig.~\ref{fig4}(d)], which
has a complicated strong shell structure including subshell effects
\cite{MSIS12}.
Thus, we also present 
the results for two deformed nuclei,
$^{166}$Ho and $^{230}$Th,
from both sides of the desired heavy particle-number
    interval
    $A\approx 140-240$.

In 
Fig.~\ref{fig4}, 
the results of the 
MMA approaches
(1 and 2)
are compared with 
 those of the well-known ``Bethe3''
\cite{Be36} [Eq.~(\ref{ldBethe})]
 asymptote; see also Table \ref{table-1} for these and a few
other asymptotical approaches,  the FG [Eq.~(\ref{intFGSS})] and,
with a focus on shell
effects,
GFG [Eq.~(\ref{SPMgen})].  Results for the MMA2a, the MMA2 [Eq.~(\ref{rho52})]
at the dominating 
    shell effect case  (ii)
    [$\xi^\ast \gg 1$, Eq.~(\ref{par}), 
     in the
    saddle point $\beta=\beta^\ast$ for 
large excitation energies $U$], and for those
with 
realistic relative shell correction $\mathcal{E}_{\rm sh}$ \cite{MSIS12}, are shown
versus the results of a  small 
    shell effects approach MMA1 (i),  Eq.~(\ref{rho32})
        ($\xi^\ast \ll 1$ at $\beta=\beta^\ast$).
     For a very small 
        value of $\mathcal{E}_{\rm sh}$, but still within the values of the case (ii),
     Eq.~(\ref{rho52})
with (\ref{rhobar52TF}) (in particular,
large $\xi^\ast$), 
     we have the approach named 
    MMA2b.
Results for the MMA2b approach are 
also shown in Fig.~\ref{fig4}.
 Results of calculations within the full SPM GFG asymptotical approach, 
Eq.~(\ref{SPMgen}), and within the popular FG approximation, Eq.~(\ref{intFGSS}), 
which are in good agreement with the
standard Bethe3 approximation, 
are only presented 
in Table \ref{table-1}.
      For finite realistic values of
      $\mathcal{E}_{\rm sh}$,  the results of the MMA2a approach 
      are
      closer to those of the MMA1 approach.
      Therefore, since the 
      MMA2b approach,
     Eqs.~(\ref{rho52}) with (\ref{rhobar52TF}), is the
     limit of the MMA2 
         one at a very
          small $\mathcal{E}_{\rm sh}$  within the case (ii),
           we conclude that the MMA2 approach
      is a more general  shell-structure
      MMA formulation of the statistical level-density problem.

In 
all panels
of 
Fig.~\ref{fig4}, one can see the divergence
of the level densities of the Bethe formula
[also, the  FG, Eq.~(\ref{intFGSS}),
    and the  GFG, Eqs.~(\ref{SPMgen}) and 
(\ref{rhoasgen})], 
near the zero excitation energy,
$U\rightarrow 0$. This is, obviously, in contrast to any MMAs,
 combinatorics
expression 
(\ref{den0gen})
in 
the limit of zero excitation energy;
see 
Eqs.~(\ref{denbes}), (\ref{rho32}),
and
(\ref{rho52}).
The MMA1 results are close to the Bethe, FG and GFG approaches
everywhere, for all presented nuclei. The reason is that
their differences are
essential only for extremely small excitation energies $U$ where
MMA1 is finite while 
other (Bethe,  FG and GFG) 
approaches 
are divergent. 
However, there are almost no excited states
in the range of their differences 
in the nuclei under
consideration.

     The results of the MMA2b approach [the same as MMA2 approach,
    Eq.~(\ref{rho52}) but with Eq.~(\ref{rhobar52TF}) for
        the coefficient $\overline{\rho}_{5/2}$, at
relatively very small shell correction,
$\mathcal{E}_{\rm sh}$] within the case (ii), for
$^{166}$Ho [see Fig.~\ref{fig4}(b)] with $\sigma $ of the order of one
 are in significantly better agreement with experimental
data
as compared 
to the results of all other approaches
(for the same nucleus). 
The MMA1
[Eq.~(\ref{rho32})],
Bethe  [Eq.~(\ref{ldBethe})], 
 FG
[Eq.~(\ref{intFGSS})], and full
SPM GFG [Eq.~(\ref{SPMgen})] approaches
are characterized by values of $\sigma\gg 1 $, 
 which are
    largely of the 
    order of
     10 (see Table \ref{table-1}).
    In contrast to the
    $^{166}$Ho excitation energy spectrum with 
   many very LESs below
    about 1 MeV, for
$^{144}$Sm (a) and 
    $^{208}$Pb (c) one finds 
    no such states.
            For
       the MMA2b [MMA2 for very small $\mathcal{E}_{\rm sh}$, but within the (ii)]
    approach we have larger values of $\sigma$,  $\sigma \gg 1$ for $^{144,148}$Sm and 
     little larger for
     $^{208}$Pb, versus 
     those of other approximations. In
    particular, for  MMA1 (i), and the
    other asymptotic approaches of
    Bethe,  FG, and GFG, 
   one finds almost the same 
$\sigma $ of the order of one, that is in
better agreement 
with data \cite{ENSDFdatabase,HJ16}.
 We obtain basically
the same for 
MMA2a (ii) with realistic values of
$\mathcal{E}_{\rm sh}$. 
Notice 
that for  $^{144,148}$Sm and 
     $^{208}$Pb nuclei, the MMA2a [Eq.~(\ref{rho52})]  at realistic $\mathcal{E}_{\rm sh}$
 is 
close
to the MMA1 (i), Bethe,  FG, and GFG approaches.
The MMA1 and MMA2a (at
realistic values of $\mathcal{E}_{\rm sh}$) as well as
Bethe,  FG and GFG approaches 
are
obviously in much better agreement with experimental data \cite{ENSDFdatabase}
for $^{144}$Sm (or $^{148}$Sm) and $^{208}$Pb
[Fig.~ \ref{fig4}(a) and (c)],
for which one has the opposite
situation: very small states number in the LES range.

We note that the results of 
the
MMA1 and
MMA2a 
with shifted excitation energies $U\rightarrow
U_{\rm eff}=U-E_{\rm cond}>0$ 
by constant condensation energies $E_{\rm cond}\approx
1.1$ and $2.2$ MeV, shown by arrows in Fig.~\ref{fig4} 
for $^{144}$Sm 
and $^{208}$Pb, respectively,  may indicate
 the pairing phase transition effect  because of disappearance of the pairing
correlations \cite{Ig83,So90,SC19}.
With increasing $U$, one can see a sharp jump 
in the level density for the double magic $^{208}$Pb nucleus 
within
the shown spectrum range.
In $^{144}$Sm, one finds such a phase transition a little above the
presented 
range of the excitation energies.
This effect could be related to the pairing phase
transition\footnote{For temperature
dependence of the pairing gap in the simplest BCS theory, one can evaluate
$\Delta(T)-\Delta_0=-\sqrt{2\pi\Delta_0T}\exp(-\Delta_0/T)$, where
$\Delta_0 \approx 12/A^{1/2}$ MeV at $T=0$; see
Refs.~\cite{SY63,Mo72,Ig83,So90,Sv06,SC19}.
Therefore, for disappearance of pairing gap, the critical temperature,
$T_{cr}=\gamma \Delta_0/\pi$, where $\gamma$ is defined by the
Euler constant, $\ln \gamma=0.577...$.
Evaluating the condensation energy, $E_{\rm cond}=g \Delta_0^2/4=
3A\Delta_0^2/(2\pi^2K)$, one arrives at the effective excitation energy,
    $U_{\rm eff}=U-E_{\rm cond}$.}
  near the critical temperature $T_{\rm cr}=0.47$ MeV in $^{208}$Pb
  ($0.57$ MeV in $^{144}$Sm), i.e. at the critical effective excitation
  energy, $U_{\rm eff}=U-E_{\rm cond}\approx 3.3$ MeV ($4.1 $ MeV in $^{144}$Sm),
  resulting in
  a level density jump. 
  These simple estimates 
  show a qualitative agreement,  by order of magnitude,
  with the condensation
  energy, $E_{\rm cond}\approx 1$ MeV. This procedure is a
      self-consistent calculation. 
      Starting from a value of the
      condensation energy, $E_{\rm cond}$,
      one can obtain the inverse level density parameter $K$. 
      Then, 
      one evaluates a  new $E_{\rm cond}$ 
     and  reiterates till convergence in the values of
      $K$ and $E_{\rm cond}$ is achieved,
      at least in order of the magnitudes. This can be realized
      for the MMA1 for $^{144}$Sm
      and MMA2a for $^{208}$Pb; see Table \ref{table-1} and
      Fig.~\ref{fig4}(a) and (c).  
      The phase transition jump is well seen
      in the
 plot (c) but is not seen in plot (a) being above the excitation
 energy range, at both
      the effective excitation energies $U_{\rm eff}$ mentioned above.

     One of the reasons  for the
     exclusive properties of $^{166}$Ho
 [Fig.~\ref{fig4}(b)]
    as compared to
    both  $^{144}$Sm (a) and $^{208}$Pb (c) 
        might be assumed to be        
        the nature
    of the excitation energy
    in these nuclei.
    Our MMA (i) or (ii) approaches could clarify the excitation nature
    [see Sec.~\ref{subsec-I} and Appendix \ref{appB} for the rotational
    contribution which can be included
    in $E_0$ of Eq.~(\ref{OmadF}) 
   as done in Eq.~(\ref{Omad})]. 
    Since
    the  results of the MMA2b (ii) approach are in much better
    agreement with experimental data than those of the MMA1 (i)  approach
    for $^{166}$Ho,
    one could presumably 
    conclude
    that for $^{166}$Ho one finds more
    clear thermal excitations, $U \gg E_{\rm rot}$, Eq.~(\ref{condU}),  for LESs.
    For  $^{144}$Sm  and $^{208}$Pb
    one observes more regular
    (large spins owing to the alignment)
            excitation contributions for dominating rotational energy
        $E_{\rm rot}$, Eq.~(\ref{condI2}); see Ref.~\cite{KM79}.
        The latter effect is
    much less pronounced in
    $^{208}$Pb than in  $^{144}$Sm,
    but all the inverse level density parameters $K$ are significant for states
    below neutron
    resonances; see Table \ref{table-1}. However, taking into account
        the pairing effects, even qualitatively,
        the thermal contribution (ii) is also important for
    $^{208}$Pb while the regular nonthermal 
        motions might be dominating in  $^{144}$Sm.
    In any case, the shell effects
    are important, 
    especially for the (ii) case which
    does not even exist
     without 
    taking them into account.

For  $^{230}$Th [Fig.~ \ref{fig4}(d)],
 one 
has the experimental LESs data
in the middle of two limiting cases  
MMA1 (i) and MMA2b (ii).
This agrees also with an intermediate number of very LESs in this nucleus. 
As shown in Fig.~\ref{fig4}(d) and Table \ref{table-1}, the MMA2a approach
at 
realistic values of $\mathcal{E}_{\rm sh}$ is in good agreement with
the data.
    The shell structure is, of course, not so strong in $^{230}$Th as compared
    to that of the  double
    magic 
    nucleus,  $^{208}$Pb,
but it is of the same order as in other presented nuclei.  Also notice 
that,
in contrast to the spherical nuclei in Figs.~\ref{fig4} (a) and (c), the nuclei
$^{166}$Ho (b) and  $^{230}$Th (d) 
are significantly deformed, which 
is also important, in particular, because of their large angular momenta of
the LES excitation spectrum states.

We do not use 
free empiric parameters
of the BSFG, spin cutoff FG,
 and  
empiric 
CTM  approaches
\cite{EB09}. 
As an advantage, one has
 only 
 the
parameter $K$
with 
the physical meaning of the inverse level density parameter.
The variations 
in $K$ are related, e.g., to those of the mean field parameters
through Eq.~(\ref{entrFG}).
All the densities $\rho(E,A)$ compared in Fig.~\ref{fig4} and  Table \ref{table-1}
do not depend 
on the cutoff spin factor
and moment of inertia 
because of summation (integrations) over all spins 
(however, with accounting for the degeneracy $2I+1$ factor).

In line with the results of Ref.~\cite{ZS16},
the obtained values   of
$K$  for the MMA2 approach can be  
essentially different from the MMA1 ones and those 
(e.g., FG) found, mainly, for the neutron resonances (NRs). 
However, the 
level densities with the excitation energy shifted 
 by constant
       condensation energies, due  
        to pairing,
        for $^{208}$Pb (c) and
        $^{144}$Sm (a) in Fig.~\ref{fig4}, 
         notably improve the comparison
with data
\cite{ENSDFdatabase}.  These densities correspond to 
inverse level-density
 parameters
$K$, smaller even than those obtained
in the FG
approach which agreed with NR data.
 We note that for the MMA1 approach one finds values of $K$ which are
of the same order as
those of the Bethe,  FG and GFG approaches.
These values of $K$ are mostly 
close to the NR values in order of 
magnitude. For the FG approach, Eq.~(\ref{intFGSS}),
in accordance with 
 its nondirect  derivation  through the spin-dependent
    level density $\rho(E,A,I)$,  Eq.~(\ref{rhoIexp2})
    (Sec.~\ref{subsec-I}),
it is obviously because the neutron resonances
occur at large
excitation energies $U$ and
small spins; see Eqs.~(\ref{condU}) and (\ref{Iexp}).
Large deformations, neutron-proton asymmetry,
spin dependence for 
deformed nuclei, and 
pairing correlations 
  \cite{Er60,Ig83,So90,AB00,AB03,ZK18,Ze19} 
 in rare earth and
      actinide nuclei should be also
      taken into account to improve the comparison with experimental data.

      \section{Conclusions}
      \l{sec-concl}
    
We derived the statistical level density $\rho(S)$ as function of the
entropy $S$
within the micro-macroscopic 
approximation (MMA) using the mixed micro- and grand-canonical ensembles beyond
the standard saddle point method of the Fermi gas model.
 The obtained level density
can be applied for small and relatively large entropies
$S$ or excitation energies
$U$ of a nucleus. For a large entropy (excitation energy), one obtains
the 
exponential asymptote of the standard SPM  Fermi gas model, 
 but with 
significant powers of $1/S$ 
corrections.
For small $S$ one finds the usual finite combinatorics
expansion in  powers of $S^2$. Functionally, the MMA at
linear approximation in $S^2 \propto U$
    expansion, at small excitation energies $U$,
    coincides with the empiric
    constant ``temperature'' model
     except it is obtained without using 
    free fitting parameters. Thus,
    MMA unifies the commonly accepted 
    Fermi gas approximation with the empiric CTM
    for large and small entropies $S$, respectively, in line 
  with  the suggestions in Refs.~\cite{GC65,ZK18}.
The MMA 
    clearly manifests an 
advantage
over the standard full SPM approaches
at low excitation energies, because 
 it does not diverge
in the limit of
small
excitation energies, in contrast to
every full SPM  approaches, e.g., 
Bethe asymptote and FG 
asymptote.
Another advantage applies when 
nuclei 
have 
many more states in the very low energy state
range.
The values of the inverse
level density
parameter $K$ 
 were compared with those of experimental data for LESs below
neutron resonances (NRs) in 
spectra of 
several nuclei. 
The MMA 
results with only one physical parameter in the least mean-square fit,
the inverse level density parameter $K$, 
 were usually 
better with larger number of the extremely low energy states,
certainly much better
than for the results with the FG model in this case.  The 
MMA values
of the inverse level density parameter $K$
for LESs 
can be significantly
 different from those of the neutron resonances within the FG model.

We found significant shell effects in the MMA level density for the
nuclear LES range
within the semiclassical periodic orbit theory.
In particular, we generalized the known SPM results for the level density in
terms of the full SPM GFG approximation accounting for the shell effects
using the POT.
Exponential disappearance of shell effects with increasing temperature was
analytically studied within the POT for
the level density. 
 Shifts 
in the entropy $S$ and in the inverse
level density parameter $K$
due to the shell effects 
were also obtained and given in the explicit analytical forms.
The shifts occur
at
temperatures 
much 
lower than the chemical potential,
near the NR excitation energies.

 Simple estimates of pairing effects
in spherical magic nuclei,
by pairing condensation energy to the excitation energies shift,
significantly improve
the comparison with experimental data.
  Pairing correlations 
essentially influence the level density parameters at low
excitation energies.
We found 
an attractive 
 description of the
well-known jump
in the level density within our MMA approach using 
the pairing phase transition.
 Other analytical reasons for the excitation energy shifts
    in the BSFG model are found by also
    using a
more accurate expansion of the modified Bessel expression for the MMA level density
at large entropies $S$, 
    taking into account high order terms in $1/S$. 
 This is important
in both the LES and NR regions, especially for LESs.
We presented a 
reasonable description of the LES experimental data for the statistical averaged level density
obtained by the sampling method
within the MMA with the help of the semiclassical POT. 
We 
      have emphasized the
importance of the shell and pairing effects in these
calculations.   
We obtained 
values of the inverse level
density parameter $K$ for the LES range which are essentially different
from those 
of NRs.
These results are basically extended to the level density dependence on the 
    spin variables
for   nuclear 
rotations
around the symmetry axis of the mean field 
 due to alignment of the
individual nucleon angular momenta along the symmetry axis.

 Our approach can be applied  to 
statistical analysis of
experimental
data on collective nuclear states.
As the semiclassical POT MMA is better with larger particle number
    in a Fermi system, one can also apply this method 
    to study
    metallic clusters and quantum dots
    in terms of the statistical level density,  and  to 
    problems
    in nuclear astrophysics.
  The neutron-proton asymmetry, 
large nuclear angular momenta  and
deformation for collective rotations,
additional consequences of pairing correlations, as well as  other perspectives, will be 
taken into account in a future work in order to improve the comparison of the
theoretical 
results with
experimental data on the level density parameter significantly,  in particular  below the
neutron resonances.

%%%%%%%%    HEAD LINE FOR ACKNOWLEDGEMENT    %%%%%%%%%%%%%%

\bigskip
\centerline{{\bf Acknowledgment}}
\medskip

The authors gratefully acknowledge Y.\ Alhassid,
D.\ Bucurescu, R.K.\ Bhaduri, M.\ Brack,
 A.N.\ Gorbachenko, and V.A.\ Plujko
for creative discussions.
 This work was supported in part by the  budget program
``Support for the development
of priority areas of scientific researches,'' a the project of the
Academy of Sciences of Ukraine (Code 6541230, No 0120U100434).
S.\ S.\ is partially supported by the US
Department of Energy under Grant No. DE-FG03-93ER-40773.

\vspace{0.5cm}

\appendix
\renewcommand{\theequation}{A.\arabic{equation}}
\renewcommand{\thesubsection}{A\arabic{subsection}}
  \setcounter{equation}{0}

\vspace{0.2cm}
\section{
The semiclassical POT}
\l{appA}

So far we did not specify the model for the mean field.
 For nuclear rotation, it can be associated with
alignment of the individual angular momenta of nucleons called a
``classical rotation'' in Ref.~\cite{KM79}: 
rotation parallel to the symmetry axis $Oz$, in contrast to 
the  collective rotation perpendicular to the $Oz$ axis \cite{GM21}.

In particular, 
in the
case of the ``parallel'' rotation,
one has for a spherically and 
axially symmetric  potential
the explicit
partition function expression: 
\bea\l{parfun}
&\ln \mathcal{Z}= 
\sum\limits_{i}\ln\left\{1 +
\exp\left[\beta\left(\lambda - 
    \varepsilon_i+\hbar \omega m_i\right)\right]\right\}\nonumber\\
&\approx \int\limits_0^{\infty}\d \varepsilon \int\limits_0^{\infty}\mbox{d} m~g(\varepsilon,m)
\ln\left\{1+\right.\nonumber\\
&+\left.\exp\left[\beta\left(\lambda -
    \varepsilon+\hbar \omega m\right)\right]\right\}~.
\eea
Here, $\varepsilon_i$ and $m_i$ are the
s.p. energies and projections of the angular momentum on the
symmetry axis $Oz$ of the quantum states in the mean field. In the transformation
from the sum to an
integral, we introduced  the s.p. level density $g(\varepsilon,m)$ as a sum of
the smooth and oscillating shell 
components,
\be\l{spdenm}
g(\varepsilon,m)\cong \tilde{g}(\varepsilon,m)+
\delta g_{\rm scl}(\varepsilon,m)~.
\ee
 The Strutinsky smoothed s.p. level density $\tilde{g}$
can be well approximated by
the ETF level density $g^{}_{\rm \tt{ETF}}$, $\tilde{g}\approx g^{}_{\rm \tt{ETF}}$.
For the spherical case, the s.p. level density in the TF
approximation is given by \cite{Be48}
\be\l{tildeg}
\tilde{g}\approx g^{}_{\rm \tt{TF}}=\frac{\mu d_s}{\pi\hbar}\int\limits_{|m|}^{\ell_0^{}}\d \ell
\int\limits_{r_{\rm min}}^{r_{\rm max}}
\d r \left[2 \mu\left(\varepsilon - V(r)\right)-\hbar^2l^2/r^2\right]^{-1}~,
\ee
where $\mu$ is the nucleon mass, $d_s$ is the spin (spin-isospin) degeneracy,
$\ell_0$ is the maximum of a  possible
angular momentum of 
nucleon with energy  $\varepsilon$
in a spherical potential well $V(r)$, and $~r_{\rm min}$
and $r_{\rm max}$ are the turning points. For the oscillating component
$\delta g_{\rm scl}(\varepsilon,m)$ of the level density $g(\varepsilon,m)$,
Eq.~(\ref{spdenm}), we use, in the spherical case,
the following
semiclassical expression \cite{KM79}
derived in Ref.~\cite{MK78}:
\be\l{goscemsph}
\delta g_{\rm scl}(\varepsilon,m)=\sum^{}_{\rm PO}\frac{1}{2 \ell_{\rm PO}}
\theta\left(\ell_{\rm PO}-|m|\right)~g^{}_{\rm PO}(\varepsilon)~.
\ee
The sum is taken here over the classical periodic orbits (PO) with 
angular momenta $\ell_{\rm PO}\geq |m|$. In this sum,
$g^{}_{\rm PO}(\varepsilon)$ is the partial
contribution of the PO to the oscillating part $g_{\rm scl}(\varepsilon)$
of the semiclassical level density $g(\varepsilon)$ (without limitations
on the projection
$m$ of the particle angular momentum), see 
Eq.~(\ref{spden}), with 
\be\l{goscsem}
\delta g_{\rm scl}(\varepsilon)=\sum^{}_{\rm PO}g^{}_{\rm PO}(\varepsilon)~,
\ee
where
\be\l{goscPO}
g^{}_{\rm PO}(\varepsilon)=\mathcal{A}_{\rm PO}(\varepsilon)
~\cos\left[\frac{1}{\hbar}\mathcal{S}_{\rm PO}(\varepsilon)-
\frac{\pi}{2} \mu^{}_{\rm PO}
-\phi^{}_0\right].
\ee
 Here, $\mathcal{S}_{\rm PO}(\varepsilon)$ is the classical action along the
 PO, $\mu^{}_{\rm PO}$ is the so called Maslov index determined by
the catastrophe points (turning and caustic points) along the PO, and
$\phi^{}_0$ is an additional shift of the phase coming from the dimension
of the problem and degeneracy of the POs. The amplitude
$\mathcal{A}_{\rm PO}(\varepsilon)$  in Eq.~(\ref{goscPO}) is 
a smooth function of
the energy
$\varepsilon$,  depending 
on the PO stability factors
\cite{SM76,BB03,MY11}.
      For the spherical cavity one has the famous explicitly analytical formula
      \cite{BD72,SM76,BB03}.
      The Gaussian local averaging of the level density shell
      correction 
          $\delta g^{}_{\rm scl}(\varepsilon)$ (Eq.~(\ref{goscsem}))
          over the
  s.p.\ energy spectrum
      $\varepsilon_i$ near the Fermi surface $\varepsilon^{}_F$ can be done
  analytically by using the linear expansion of
  relatively 
  smooth PO action
      integral $\mathcal{S}_{\rm PO}(\varepsilon)$ 
          near $\varepsilon^{}_F$ as function of $\varepsilon$
 with the Gaussian width parameter $\Gamma$ \cite{SM76,BB03,MY11},
\be\l{avden}
\delta g^{(\Gamma)}_{\rm scl}(\varepsilon) \cong
\sum^{}_{\rm PO}g^{}_{\rm PO}(\varepsilon)~
\exp\left[-\left(\frac{\Gamma t^{}_{\rm PO}}{2\hbar}\right)^2\right]~,
\ee
where $t^{}_{\rm PO}=\partial S_{\rm PO}/\partial \varepsilon$ is the period
of particle motion along the PO.
    All the expressions presented above, 
    except for
    Eqs.~(\ref{tildeg}) and (\ref{goscemsph}), can be applied for the axially-symmetric
    potentials, e.g.\ for the spheroidal cavity \cite{SM77,MA02,MY11} and deformed
    harmonic oscillator
    \cite{Ma78,BB03}.

Let us use now the decomposition of 
  $\Omega\equiv-\ln\mathcal{Z}/\beta$ with the corresponding variables
  within the SCM POT in terms of its smooth part, $\tilde{\Omega}\approx
  \Omega^{}_{\rm \tt{ETF}}$, and shell correction $\delta \Omega$:
  \be\l{SCMpot}
  \Omega\left(\beta,\lambda,\omega\right) 
  \cong ~\tilde{\Omega}\left(\beta,\lambda,\omega\right) 
  +
  \delta \Omega\left(\beta,\lambda,\omega\right)~.
  \ee
Using the TF approximation for 
$\tilde{g}(\varepsilon,m)$,  Eq.~(\ref{tildeg}), for a smooth TF
component $\Omega^{}_{\rm \tt{ETF}}$ of the
potential $\Omega$, Eq.~(\ref{SCMpot}), one has \cite{KM79}
\bea\l{TFpot}
&\tilde{\Omega}\approx \Omega^{}_{\rm \tt{ETF}}\left(\beta,\lambda,\omega
\right)
=-\beta^{-1} 
\int\limits_0^\infty\d\varepsilon\int\limits_{-\infty}^{\infty}\d m~
\tilde{g}(\varepsilon,m)~\nonumber\\
&\times\ln\left\{1+\exp\left[\beta\left(\lambda-
  \varepsilon+
  \hbar\omega
  ~m\right)\right]\right\}\nonumber\\
=&\tilde{E} 
-\lambda A
-\frac12\tilde{\Theta} (\lambda)~\omega^{2}
-\frac{\pi^2}{6}\tilde{g}(\lambda)\beta^{-2}~.
\eea
 The smooth 
(in the sense of the SCM 
\cite{St67,BD72})
ground-state 
energy of the nucleus is given by 
  \be\l{TFE0}
 \tilde{E}\approx E^{}_{\rm \tt{ETF}}=\int_0^{\tilde{\lambda}}
  \d \varepsilon~\varepsilon~ \tilde{g}(\varepsilon)\approx \int_0^{\lambda}
  \d \varepsilon~\varepsilon~ \tilde{g}(\varepsilon)~,
  \ee
   where
  $\tilde{g}(\varepsilon)$
  is a smooth 
  level density 
  approximately  equal to the
  ETF level density, $\tilde{g}\approx g^{}_{\rm \tt{ETF}}$. 
 The smooth  chemical potential $\tilde{\lambda}$ in the SCM
  is 
the root of equation
  $ A=
       \int_{0}^{\tilde{\lambda}}\mbox{d} \varepsilon~\tilde{g}(\varepsilon)$, and
       $\lambda \approx \tilde{\lambda}$ in the POT.
 The chemical potential $\lambda$ (or  $\tilde{\lambda}$)
  is approximately the solution of the corresponding conservation particle number 
  equation:
  \be\l{chempoteq}
    A~=\int_{0}^{\lambda} \mbox{d} \varepsilon~g(\varepsilon)~.
\ee
   The quantity $\Theta^{}_{\rm \tt{ETF}}$ in Eq.~(\ref{TFpot}) is
   the ETF (rigid-body) moment of inertia for the statistical
   equilibrium rotation,
\bea\l{rigMIpar}
&\tilde{\Theta}\approx \Theta^{}_{\rm \tt{ETF}}=
  \mu\int \d {\bf r}~\tilde{\rho}({\bf r})~(x^2+y^2)\nonumber\\
&\approx \hbar^2
\langle \widetilde{m^2}\rangle~\tilde{g}\left(\lambda\right)~,
\eea
where $\tilde{\rho}\approx \rho^{}_{\rm \tt{ETF}}({\bf r})$
is the ETF 
particle density. 
For the ``parallel'' rotation, 
$\langle \widetilde{m^2}\rangle$
is 
the smooth component of
the square
of the angular momentum
projection of nucleon $\langle m^2\rangle$. 
Here and below we neglect a small change 
in the chemical
 potential $\lambda$, 
due to the internal nuclear  thermal and rotational excitations, 
which can be approximated by the Fermi energy $\varepsilon^{}_F$,
$\lambda\approx \varepsilon^{}_F$.

 The oscillating semiclassical component
$\delta \Omega\left(\beta,\lambda,\omega\right)$ 
of the
sum (\ref{SCMpot}) 
corresponds to the oscillating part
$\delta g_{\rm scl}(\varepsilon,m)$ of the
level density (\ref{spden}) [see, e.g.,
  Eq.~(\ref{goscemsph}) for the spherical case]
\cite{SM76,KM79,MK78}.
In expanding the action $\mathcal{S}_{\rm PO}(\varepsilon)$ as function of the
s.p. energy $\varepsilon$ near the chemical potential $\lambda$ in
powers of $\varepsilon-\lambda$ up to linear term 
one can use Eqs.~(\ref{goscsem}) and (\ref{goscPO}); 
 see also Eqs.~(\ref{FESCF}), (\ref{dFESCF}), and
(\ref{dEPO0F}).
Then, integrating by parts, one obtains 
from Eqs.~(\ref{parfun}),
(\ref{SCMpot}), and (\ref{TFpot}) at
the adiabatic approximation
$\hbar \ell^{2}_F\omega 
\ll \lambda$,  
where $\hbar \ell^{}_F$ is the maximal s.p. spin at the Fermi surface, the result
\bea\l{potoscpar}
&\delta \Omega \cong \delta \Omega_{\rm scl}\left(\beta,\lambda,\omega\right)
=\delta F_{\rm scl}\left(\beta,\lambda,\omega\right)\nonumber\\
&=\delta F_{\rm scl}\left(\beta,\lambda\right) 
 -\frac{\omega^2}{6} \sum^{}_{\rm PO} F_{\rm PO}~
t^{2}_{\rm PO}~l^{2}_{\rm PO}~,
\eea
where $\delta F_{\rm scl}\left(\beta,\lambda\right)$ is the
semiclassical free-energy shell correction of a nonrotating nucleus ($\omega=0$);
see Eqs.~(\ref{FESCF})
and (\ref{dFESCF}). 
 In deriving the expressions for the
free energy shell correction
$\delta F_{\rm scl}$ 
and the potential $\delta \Omega_{\rm scl}$, the action
$\mathcal{S}_{\rm PO}(\varepsilon)$ in their integral representations over
$\varepsilon $ with the semiclassical level-density shell correction
$\delta g(\varepsilon)$, Eqs.~(\ref{goscsem}) and (\ref{goscPO}),
was expanded as function of $\varepsilon$
near the chemical potential $\lambda$. 
 Then, we integrated by parts over $\varepsilon$, as in
the semiclassical calculations of the energy shell correction
$\delta E_{\rm scl}$  \cite{SM76,BB03}. 
We used the expansion of
    $\delta \Omega(\beta,\lambda,\omega)$ over a relatively small rotation frequency $\omega$,
$\hbar \ell^{2}_F\omega/\lambda\ll 1$, up to
quadratic terms.  
 Nonadiabatic effects for large $\omega$,
    considered in 
Ref.~\cite{KM79} for the spherical case, are out
of the scope of this work.
In Eq.~(\ref{potoscpar}), the period of motion along a PO,
$t^{}_{\rm PO}(\varepsilon)=\partial S_{\rm PO}(\varepsilon)/\partial \varepsilon$,
and
the PO angular momentum of particle, $\ell^{}_{\rm PO}(\varepsilon)$,
are taken at $\varepsilon=\lambda$. 
For large excitation energies,
$\beta=\beta^{\ast}=1/T$ ($T$ is the temperature), 
one arrives from Eqs.~(\ref{FESCF}), (\ref{dFESCF}), and (\ref{potoscpar}) at
the well-known expression for the semiclassical free-energy shell correction
of the POT \cite{KM79,BB03}, $\delta F=\delta \Omega$
    (in their specific variables);
see also Ref.~\cite{Ra97} for the
magnetic-susceptibility
shell corrections. These shell corrections decrease exponentially with
increasing temperature $T$. For the opposite limit to the yrast line
(zero excitation energy $U$, $\beta^{-1}\sim T \rightarrow 0$), one obtains from
$\delta \Omega$, Eq.~(\ref{potoscpar}), the well-known
POT approximation \cite{SM76,BB03} to the energy shell correction
$\delta E$, modified however by the frequency $\omega $ dependence.

The POT shell effect component of the free energy, $\delta F_{\rm scl}$, Eqs.~(\ref{FESCF})
and (\ref{dFESCF}),
is related in the nonthermal and nonrotational limit to the energy
    shell correction of a cold nucleus,
$\delta E_{\rm scl}$
\cite{SM76,BB03,MY11,MG17}:
\be\l{escscl}
\delta E_{\rm scl} = \sum_{\rm PO}E_{\rm PO}= \sum_{\rm PO}\frac{\hbar^2}{t_{\rm PO}^2}\,
g^{}_{\rm PO}(\lambda) %E_{\rm PO}
~,
\ee
 where $E_{\rm PO}$ is the partial PO component [Eq.~(\ref{dEPO0F})] of
the energy shell correction
$\delta E$. 
Within the POT, $\delta E_{\rm scl}$
is determined, in turn, by 
the oscillating level density $\delta g_{\rm scl}(\lambda)$;
see Eqs.\ (\ref{goscsem}) and (\ref{goscPO}).

The chemical potential $\lambda $ can be approximated by the Fermi energy
$\vareps^{}_F$, up to small excitation-energy and rotational-frequency
corrections ($T\ll \lambda$
for the saddle point
value $T=1/\beta^\ast$ if it exists, and
$\hbar\ell^{}_F\omega/\lambda \ll 1$).
  It is determined by the particle-number conservation condition, Eq.~(\ref{conseqparA}),
   which can be written in the simple form (\ref{chempoteq}) 
      with the total POT level density
  $g(\vareps)\cong g^{}_{\rm scl}=g^{}_{\rm \tt{ETF}} +\delta g^{}_{\rm scl}$.
One now needs to solve 
equation (\ref{chempoteq}) for a given particle number $A$ to
determine the 
chemical potential $\lambda $ as
   function
  of 
  $A$, since 
$\lambda$  is needed in Eq.\  (\ref{escscl}) to 
obtain the semiclassical energy shell correction
$\delta E_{\rm scl}$. If one were to 
use in Eq.~(\ref{chempoteq}) the exact (SCM) level density
$g(\vareps)\approx g^{}_{\rm SCM}=
\tilde{g}+\delta g^{}_{\Gamma}(\varepsilon)$, where
$\tilde{g}$ is the Strutinsky smooth s.p. level density,
$\tilde{g}\approx g^{}_{\rm \tt{ETF}}$, and $\delta g^{}_{\Gamma}$ is the
averaged level-density shell correction with Gaussian width $\Gamma$,
one would obtain
a steplike function of the needed
chemical potential $\lambda$ (Fermi energy $\varepsilon^{}_F$)
as a function of the
particle number $A$. 
Using the semiclassical level density
$g_{\rm scl}(\vareps)$, Eq.~(\ref{spden}),
with
$\delta g_{\rm scl}(\vareps)$ 
given by Eqs.~(\ref{goscsem})  
 and (\ref{goscPO}), similar 
discontinuities would appear. To avoid such a behavior, one can apply the
Gauss averaging, e.g., Eq.\ (\ref{avden}), on the level density
$g^{}_\Gamma(\vareps)$ in 
      Eq.~(\ref{chempoteq}) 
      or, what amounts to the same, on the quantum SCM states density with,
      however, a width 
      $\Gamma=\Gamma^{}_0$. This Gauss width
        should  be much smaller than 
        that obtained in 
        a 
    shell-correction calculation,
    $\Gamma=\Gamma_{\rm sh}$,
with 
$\Gamma^{}_0 \ll \Gamma^{}_{\rm sh} \ll D_{\rm sh}$,
where $D_{\rm sh}$ is the distance between major shells.
Because of 
a slow convergence of the PO sum in Eq.\ (\ref{goscsem}), it is,
however, more 
practical to use in Eq.~(\ref{chempoteq}) 
the SCM quantum density, 
$g(\vareps)\approx g^{}_{\rm SCM}(\varepsilon)$, averaged with $\Gamma^{}_0$
to determine the function 
$\lambda(A)$.

For a major shell structure near the Fermi energy surface,
$\varepsilon\approx \lambda$,
the POT shell 
 correction $\delta E_{\rm scl}$ [Eq.~(\ref{escscl})]
 is in
 fact approximately proportional to that of 
$\delta g_{\rm scl}(\lambda)$ 
[Eqs.\ (\ref{goscsem}) and (\ref{goscPO})].
Indeed, the rapid convergence of the PO
sum in Eqs.~(\ref{escscl}) and (\ref{dEPO0F})
is guaranteed by the 
factor in front of the density component $g^{}_{\rm PO}$,
Eq.\ (\ref{goscPO}), a factor 
which is inversely proportional to the period
time $t^{}_{\rm PO}(\lambda)$ squared along 
the PO. Therefore, only POs with 
short periods which occupy a 
significant 
phase-space volume near the Fermi surface will contribute.
These orbits are responsible for the
major shell structure, that is related to a Gaussian averaging width,
$\Gamma\approx \Gamma_{\rm sh}$, which is much larger
than the distance between neighboring s.p. states but much smaller
than the distance
$D_{\rm sh} $ between major shells near the Fermi surface.
According to the POT \cite{SM76,BB03,MY11},
the distance between major shells, $D_{\rm sh}$, is
determined by a
mean period of the  
shortest and most degenerate POs, $\langle t^{}_{\rm PO}\rangle$
\cite{SM76,BB03}:
\be\l{periode}
D_{\rm sh} \cong 
\frac{2\pi \hbar}{\langle t^{}_{\rm PO}\rangle} 
\approx \frac{\lambda}{A^{1/3}}~.
\ee
Taking the factor in front of
$g^{}_{\rm PO}$ in 
the energy shell correction
$\delta E_{\rm scl}$, Eq.~(\ref{escscl}), off the sum over the POs, one arrives at
Eq.~(\ref{dedg}) for the semiclassical energy-shell correction
\cite{SM76,SM77,MY11,MG17}. Differentiating Eq.~(\ref{escscl}) 
 using (\ref{goscPO}) with respect to $\lambda$
and keeping only the dominating terms coming from 
differentiation of the sine of the action phase argument,
$S/\hbar \sim A^{1/3}$, one finds the useful relationship 
\be\l{d2Edl2}
\frac{\partial^2\delta E_{\rm PO}}{\partial\lambda^2}\approx -\delta g^{}_{\rm PO}~.
\ee
By the same semiclassical arguments, the dominating contribution to
$g''(\lambda)$ for major shell structure is given by
\be\l{d2g}
\frac{\partial^2 g}{\partial\lambda^2}\approx
\sum^{}_{\rm PO}\frac{\partial^2\delta g_{\rm PO}}{\partial\lambda^2}
  \approx -\left(\frac{2\pi}{D_{\rm sh}}\right)^2 \delta g(\lambda)~.
\ee
Again, as in the derivation of 
 Eqs.~(\ref{dedg}) and (\ref{d2Edl2}),
for the major shell
structure, we take the averaged smooth characteristics for the main
shortest POs which occupy the largest phase-space volume off the PO sum.  

\renewcommand{\theequation}{B.\arabic{equation}}
\renewcommand{\thesubsection}{B\arabic{subsection}}
 \setcounter{equation}{0}

\section{MMA spin-dependent level density}
\l{appB}

 For 
statistical
description of the level density of a nucleus in
  terms of the conservation variables, 
the total energy $E$, nucleon number  $A$, 
and
the angular momentum projection $M$  to a space-fixed axis $Oz$,
one
can begin with
the micro-canonical expression for the level density, 
\be\l{dendef}
\rho(E,A,M)=
\sum_i~\delta(E-E_i)~\delta(A-A_i)~
\delta(M-M_i) 
~,
\ee
where $E_i$, $A_i$, and $M_i$, 
respectively, represent the  system quantum energy spectrum.
This level density
can be identically rewritten in terms of
the inverse Laplace transformation of the partition function
$\mathcal{Z}(\beta,\alpha,\kappa)$ 
over the corresponding Lagrange multipliers
$\beta,\alpha$, and $\kappa$;
see, e.g.,
Refs.~\cite{BM67,Ig83,So90}:
\bea\l{dengen}
&\rho(E,A,M) = (2 \pi i)^{-3}\int \int
\int 
\mbox{d} \beta \mbox{d} \alpha
    \mbox{d} \kappa~
    \mathcal{Z}(\beta,\alpha,\kappa) 
    \nonumber\\
   &\times \exp\left[\beta E-A\alpha-M\kappa\right]~.
    \eea
    We will 
calculate by the SPM the
integrals in this equation 
over the restricted
set of  Lagrange multipliers
$\alpha$
and $\kappa$, related to $A$ and $M$, 
respectively.
However, as in Sec.~\ref{sec-levden},
the last integral in Eq.~(\ref{dengen})
over the variable $\beta$, related to the energy $E$, will be calculated
more 
accurately beyond the SPM approach. 
The saddle points over other variables (marked by
asterisks; see below) are
determined by saddle point 
equations:
\be\l{conseq}
A=
\left(
\frac{\partial \ln \mathcal{Z}}{\partial \alpha}\right)^\ast~,~~~
M=
\left(
\frac{\partial \ln \mathcal{Z}}{\partial \kappa}\right)^\ast~.
\ee
The asterisk mean
that $\alpha=\alpha^\ast$ and $\kappa=\kappa^\ast$.
These equations can be considered also as conservation laws
for a given set of $M$ and 
$A$.   
Equations (\ref{conseq}) for the 
saddle point values $\alpha^\ast=\lambda \beta$ and
$\kappa^\ast=\hbar\omega \beta$
in terms of the chemical potential $\lambda$ and rotation frequency
$\omega$ in the case of
axially symmetric (or spherical) mean fields 
    for the ``parallel'' rotation
(Sec.~\ref{subsec-I})
can be written in more 
explicit way:
\bea\l{conseqparM}
M&=&\int\limits_0^\infty \mbox{d} \varepsilon
  \int\limits_{-\infty}^{\infty} \mbox{d} m 
m~
g(\varepsilon,m)~n(\varepsilon,m)~,\l{conseqparA}\nonumber\\
A&=& \int\limits_0^\infty \mbox{d} \varepsilon
  \int\limits_{-\infty}^{\infty} \mbox{d} m
 ~ g(\varepsilon,m)~n(\varepsilon,m)~.
\eea
Here,
$g(\varepsilon,m)$ and $n(\varepsilon,m)$ are the s.p.\ level
density [Eq.\ (\ref{spden})] and occupation
 number, 
$~n=\left\{1+ \exp\left[\beta\left(\varepsilon-\lambda-
  \hbar m~\omega\right)\right]\right\}^{-1}~,
$
respectively.  The relations shown in
 Eq.~(\ref{conseqparM})
    are equations
        for the frequency 
$\omega$ and chemical potential $\lambda$ 
        as functions of the 
         integrals of motion,  projection of the angular momentum
$M$ and particle number $A$, respectively.

The frequency $\omega$ can be eliminated with the help of the
relations (\ref{conseq}) and (\ref{parfun}) [or Eq.~(\ref{conseqparM});
see Eq.~(\ref{Eex})]. 
 The moment of inertia (MI), $\Theta$, 
given by  Eq.\ (\ref{MI}), is
decomposed in terms of the smooth [Eq.\ (\ref{rigMIpar})] and
oscillating components. For 
a spherical potential, one can specify the MI shell correction as
\be\l{MIpar}
\delta \Theta \cong \delta \Theta_{\rm scl} =
\frac13\sum^{}_{\rm PO}t^{2}_{\rm PO}l^{2}_{\rm PO}~F_{\rm PO}~,
\ee
 where
$F_{\rm PO}$ is given by Eqs.\ (\ref{dFESCF}) and (\ref{dEPO0F}).
 In 
deriving Eq.~(\ref{MIpar}) we used explicitly the spherical symmetry
of the mean field as in Eq.~(\ref{goscemsph}) for the oscillating
level density
$\delta g_{\rm scl}(\varepsilon,m)$ and Eq.~(\ref{potoscpar}) for the potential
shell correction $\delta \Omega_{\rm scl}$. 
These components for small excitation energies and
major shell-structure averaging,
$\tilde{g}^{-1} \ll \Gamma \ll D_{\rm sh}$, of $\delta g$ are much
smaller than the average rigid body value
$\tilde{\Theta}$ [Eq.\ (\ref{rigMIpar})],
$\delta \Theta/\tilde{\Theta} \approx \delta g/ 3 \tilde{g}
\approx 2 \pi^2\mathcal{E}_{\rm sh}/3A^{1/3}\ll 1$; see Eqs.~(\ref{xibdE})
and (\ref{dedg}).
In the derivations of  Eq.~(\ref{potoscpar}) 
we used the conservation
conditions for the particle number and angular momentum projection,
Eq.~(\ref{conseq}) [or Eq.~(\ref{conseqparM})]. 
In the adiabatic approximation, one
can simplify the decomposition of the potential
$\Omega$ [Eq.\ (\ref{SCMpot})] in
terms of smooth and oscillating POT components, Eqs.\ (\ref{TFpot}) and
(\ref{potoscpar}) with (\ref{MIpar}):
\be\l{Omad}
\Omega \approx
E_0-\frac{a}{\beta^2}-\lambda A-\frac12~\Theta~\omega^2~.
\ee
This equation, which is  valid for arbitrary axially symmetric potential,
contains shell effects through the ground-state energy $E^{}_0$,
the level density parameter $a$, Eq.\ (\ref{a0par}),
and MI, Eq.\ (\ref{MI}).

 Similarly as in Eq.~(\ref{Sexp}), expanding now
$\ln \mathcal{Z}(\beta,\alpha,\kappa)$ in Eq.~(\ref{dengen}) over the variables
$\alpha$ and $\kappa$
for arbitrary
$\beta$ near the saddle points $\alpha=\alpha^{\ast}$ and $\kappa=\kappa^\ast$,
one can use Eq.~(\ref{conseq}) for the
 saddle points. 
Performing, then, the SPM Gaussian integrations over $\alpha $ and $\kappa$,
one finds 
  \bea\l{den1}
  &\rho(E,A,M)=
  (2\pi)^{-2} i^{-1}
  \int 
  \beta\d \beta
  \left[\mathcal{J}\left(
      \frac{\partial \Omega}{\partial \lambda},
      \frac{\partial \Omega}{\partial \hbar\omega};
      \lambda,\hbar\omega 
      \right)\right]^{-1/2}\nonumber\\
    &\times\!\exp\left[\beta\!\left(E-\!
      \Omega \!-\!\lambda A
      \!-\! \hbar\omega M  
      \right)
    \right]~.
  \eea
     Here, 
   $\lambda\equiv \alpha^\ast/\beta$,
  $\omega\equiv
  \kappa^\ast/\hbar \beta$, and
  $\mathcal{J}$ is the two-dimensional Jacobian
  for the
  transformation
  between the 
  two shown sets of variables.
    Finally, 
  at the saddle point of Eq.~(\ref{conseq}), 
  one can recognize
  the entropy in the exponent argument:
  \be\l{entr}
  S=\beta \left[E-\Omega(\beta,\lambda,\omega) -\lambda A -\hbar\omega M
  \right]~;
  \ee
see  Sec.\ \ref{sec-levden} 
for more 
explicit similar derivations.
      It was
      convenient also to
  introduce, instead
  of the partition function $\mathcal{Z}$ , the
  potential
 \be\l{GCEpot}
  \Omega\left(\beta,\lambda,\omega\right)= 
  - \ln \mathcal{Z}\left(
  \beta,\lambda \beta,\hbar \omega \beta\right)/\beta 
  \ee
  for any value of the integration variable $\beta$ ($\alpha=\lambda \beta$ and
  $\kappa=\hbar \omega \beta$).
    It is the well-known
    potential of the grand canonical ensemble 
    when taken at all the saddle points as
  $\Omega^\ast=
  \Omega\left(\beta^{\ast},\lambda^\ast,\omega^\ast\right)$, 
  where $\beta^\ast=1/T$ with $T$ being the system temperature,
   which, if it exists, can be defined using, $\lambda^\ast=\alpha^\ast T$,
  $\omega^\ast=\kappa^\ast T/\hbar$. %, as usual;
  We have also
  $E=\Omega^\ast+\left(\beta\partial \Omega/\partial\beta\right)^\ast +\lambda^\ast A$.
  Note that within the grand canonical ensemble, 
  the quantities $\lambda^\ast$ and $\omega^\ast$ are the standard
 chemical potential and 
       rotational frequency, respectively.
Below we consider $\lambda=\alpha^\ast/\beta$
 and $\omega=\kappa^\ast/\beta\hbar$ 
 (for any value of $\beta$)
 as the generalized chemical
potential and rotational frequency.

  The potential
  $\Omega(\beta, \lambda, \omega)$, Eq.~(\ref{Eex}),
   contains 
    two
       contributions: the thermal 
      intrinsic excitation energy, $U(\beta^\ast)=aT^2$,
      related to the entropy production, and 
      the rotational excitation energy, $E_{\rm rot}(\omega)=\Theta \omega^2/2$.
    Assuming a small thermal excitation energy, $U\propto 1/\beta^2$
    (i.e., $aT^2$
    in the asymptotically large
  excitation energy limit), with respect to rotational ones,
  $E_{\rm rot}$ (i.e., $\Theta \omega^2/2$ in the adiabatic
  approximation) but large as
  compared to a mean distance between neighbor level energies for validness of the statistical
  and semiclassical arguments, one
  writes, at $\beta \approx \beta^\ast$,
\be\l{condI2}
\frac{1}{\tilde{g}} \siml U
\approx \frac{a}{\beta^2} \ll \frac12 \Theta \omega^2~.
  \ee
The level density parameter
  $a$ is given by Eq.\ (\ref{denpar}) modified, however, by the rotational $\omega^2$ corrections:
\be\l{a0par}
a \approx \frac{\pi^2}{6}\left[g\left(\lambda\right)
+\frac{\omega^2}{6} \sum^{}_{\rm PO}g^{}_{\rm PO}\left(\lambda\right)
~t^{2}_{\rm PO}~l^{2}_{\rm PO}\right]~.
\ee
 The second term in the square brackets is  explicitly
    presented for the spherical potential.
 Note that the condition
 (\ref{condI2}) is satisfied 
  for smaller nuclear excitation energies
$U \siml 3$ MeV for typical rotational excitation energies
$\hbar \omega \siml 1$ MeV;
cf. Eq.\ (\ref{condU}).
The same 
 limit $U \simg 1/\tilde{g}(\lambda)$
 in Eqs.~(\ref{denbes}), (\ref{condU}), and the left-hand side of Eq.~(\ref{condI2})
is due to the
fact that,
in the calculation of the
quantity
$\Omega\left(\beta,\lambda,\omega\right)$, Eqs.\ 
(\ref{GCEpot}) and (\ref{parfun}), the sum over the s.p.\ states
was approximately replaced by the integral, and the continuous s.p.
level-density 
 approximation   for
$g(\varepsilon,m)$, Eqs.~(\ref{spdenm}) -- (\ref{goscPO}),
was
used. 
 In Eq.~(\ref{condI2}), for a typical rotation energy
$\hbar \omega \siml 0.1~$MeV,
one has $~0.2~\siml U \siml 3~\mbox{MeV}$  ($\lambda \approx 40$MeV).

 Under the (i) condition (\ref{condI2}) (see also Sec.~\ref{subsec-32}), 
one takes the two-dimensional
    Jacobian $\mathcal{J}$, Eq.~(\ref{Jac1F}),
    $\mathcal{J}\approx \tilde{\mathcal{J}}$, as a smooth quantity,
off the integral over $\beta$
 in Eq.~(\ref{den1}).
    Then, in the calculations of this integral,  
    we used the transformation of the variables,
$\beta=1/\tau$,
to arrive
at the integral representation 
  for the modified Bessel
  functions $I_\nu$ of the order of $\nu$ 
   (e.g., $\nu=2$). This representation
  is 
  the well-known inverse Laplace transformation \cite{AS64},
  \bea\l{Laplace}
  &\frac{1}{2\pi i}\int_{c -i \infty}^{c +i \infty}\d \tau~\tau^{-\nu-1}
  \exp\left(x \tau +y/\tau\right)\nonumber\\
  &=
  \left(\frac{x}{y}\right)^{\nu/2}I_{\nu}\left(2\sqrt{xy}\right),
  \qquad \nu >-1~,
  \eea
  where 
  $I_\nu(z)$ 
  is the same modified Bessel function of the order
  of $\nu$ as used in Eqs.\ (\ref{denbes}) and (\ref{denbes1}). In these transformations we
  assumed that the integrand
  in 
Eq.\ (\ref{den1})  is an analytical function of the integration variable 
$\tau=1/\beta$  on the right of the imaginary axis ($c>0$). This 
means that there are no
  equilibrium states (poles) for
  the excitation energy $U>0$. 
  Notice that the Jacobian
  $\mathcal{J}$ can be also
  taken off the integral over $\beta$ 
  at $\beta=\beta^{\ast}$ within the full SPM
  if the saddle point $\beta^{\ast}$ exists; see Ref.\ \cite{BM67} where 
      the assumption of constant s.p. level density near the
      Fermi surface was used.
       In the following derivations, we 
     will  neglect small thermal and rotational corrections to the
  chemical potential $\lambda$
  as compared to the Fermi energy $\varepsilon^{}_F$.
   Excitation energies of the approximate condition, Eq.~(\ref{condU}),
   should also be
   smaller than a
  distance between major shells, $D_{\rm sh}$,
  Eq.~(\ref{periode}), 
  in the adiabatic approximation
  for 
  rotational excitations. At the same time, we neglect  the 
  oscillating $\beta $
  dependence of
  the Jacobian, $\delta \mathcal{J}$ (Jacobian 
  subscript is $\infty$ in Ref.~\cite{KM79}),
  under
  the 
  condition of case (i) [see Eq.~(\ref{condI2}) and Sec.~\ref{subsec-I} 
      for the typical rotational
      energy $\hbar \omega \siml 0.1$ MeV].  Thus, 
          one finally arrives at Eq.~(\ref{denbes1})  for $\nu=2$ in the case (i).
          For the coefficient $\overline{\rho}_\nu$ in the case 
          (i) but for arbitrary
  $\nu$, one finds
\be\l{barrhon}
\overline{\rho}_{\nu}=
\frac{2 a^{\nu}}{
  \pi^{\nu-1}\left\vert\tilde{\mathcal{J}}^{(2\nu-2)}\right\vert^{1/2}}~. 
\ee
 The superscript $2\nu-2$ of the smooth part of the Jacobian,
    $\tilde{\mathcal{J}}^{(2\nu-2)}$, 
    Eq.~(\ref{Jac1F}),
provides the number of the integrals of motion 
beyond 1 (energy $E$).
    In the considered case of $n=3$ integrals
    of motion, one has
    $\nu=(n+1)/2=2$, 
        and the corresponding smooth
            Jacobian is given by 
        $\tilde{\mathcal{J}}^{(2)}\approx g^{}_{\rm \tt{ETF}}(\lambda)\Theta/\hbar^2$.

Note that 
the expressions (\ref{denbes1}) and (\ref{barrhon}), for the case (i), are presented
  in a general 
  form for axially symmetric potentials and
  arbitrary number of integrals of motion $n$. They are valid under
the condition
(\ref{condI2}), e.g., $n=3$ and $\nu=(n+1)/2=2$
in this appendix
and 
the same as in Ref.~\cite{KM79}.
For the specific case $n=2$,  the case  (i) ($\nu=3/2$) in Sec.~\ref{subsec-32},
one obtains Eq.~(\ref{rho32}), with 
Eq.~(\ref{rhoGEN32})
  for the constant $\overline{\rho}_{3/2}$, and  
 its Bethe  asymptote
(\ref{ldBethe}).

In the opposite case (ii) (Sec.~\ref{subsec-52}) 
for a
small rotational energy $E_{\rm rot}$
    with respect to the thermal excitations $U$, $E_{\rm rot}\ll U$
    [opposite to the condition (\ref{condI2})], 
for the Jacobian $\mathcal{J}$ in the integrand of Eq.~(\ref{den1}),
up to  shell corrections,
    one obtains approximately from Eqs.\ (\ref{SCMpot}), 
    (\ref{potoscpar}),
    and taking finally Eq.~(\ref{TFpot}) for
    $\Omega \approx \tilde{\Omega}$,
     the expression
\be\l{Jacsph}
\mathcal{J}=\mathcal{J}\left(
      \frac{\partial \Omega}{\partial \lambda},
      \frac{\partial \Omega}{\partial \hbar\omega};
      \lambda,\hbar\omega 
      \right) 
\approx \frac{2\,a\, \tilde{\Theta}}{\hbar^2\lambda^2 \beta^2}~.
\ee
 This Jacobian 
was simplified by
    expanding the $\beta $ depending
    factor in Eq.\ (\ref{dFESCF}) over the variable $x^{}_{\rm PO}\propto 1/\beta$
    under the condition of smallness of the $x^{4}_{\rm PO}\propto 1/\beta^4$ term;
    see Eq.~(\ref{condU}). The shell corrections in the Jacobian calculations
    in the case (ii) 
   were  neglected finally, in Eq.~(\ref{Jacsph}), as compared to the smooth (E)TF part
   $\tilde{\mathcal{J}}$
    [see similar derivations around Eq.~(\ref{rhobar52TF})]. 
       As a result of the integration over $\beta$ in Eq.~(\ref{den1})
    with Eq.~(\ref{Jacsph})
    for the Jacobian $\mathcal{J}$ and 
    the help of Eq.~(\ref{Laplace}) 
    (after transformation of the integration variables,
$\beta=1/\tau$), 
    one obtains finally the same Eq.~(\ref{denbes1}) but with
    $\nu=3$.
    The coefficient $\overline{\rho}^{}_{3}$ is given by  Eq.~(\ref{conM3}) at
    $\Theta \approx \tilde{\Theta}$.

   For large and small entropy $S$,
      one obtains  from Eq.~(\ref{denbes1}), with the
      help of Eqs.~(\ref{rhoasgen}) and (\ref{den0gen}), the asymptotic
      Fermi gas
      (at zero order in $1/S$) and
       combinatorics expressions (in powers of $S^2\propto U$),
      respectively.      
  At
  small entropy, $S \ll 1$, one obtains from Eq.~(\ref{denbes1})
  [with Eq.~(\ref{den0gen})]
 the combinatorics power expansion
 starting from a 
 constant,
that is finite in the limit $S\ll 1$. 
This expansion 
in powers of 
    $S^2 \propto U$ 
    is the same 
    as that of the empiric CTM
used often for the level density calculations  at
small excitation energies $U$ \cite{GC65,ZS16}, but
here it is obtained without free parameters.

\renewcommand{\theequation}{C.\arabic{equation}}
\renewcommand{\thesubsection}{C\arabic{subsection}}
  \setcounter{equation}{0}

\vspace{0.2cm}
\section{
Full SPM for a general Fermi gas (GFG) asymptote}
\l{appC}

Taking 
the integral (\ref{rhoE1F}) over $\beta $ by the standard
SPM,
one can expand, up to second-order terms, the exponent argument $S(\beta)=\beta U+a/\beta$
near the 
 saddle point $\beta=\beta^\ast$,
\be\l{expbeta}
S(\beta)=\beta^\ast U+a/\beta^\ast + \frac12
\left(\frac{2a}{\beta^3}\right)^\ast(\beta-\beta^\ast)^2.
\ee
The first derivative disappears because of the SPM condition, 
\be\l{spmcondbeta}
\left(\frac{\partial S}{\partial \beta}\right)^\ast\equiv U-
\frac{a}{(\beta^{\ast})^2}=0~,
\ee
from which one finds the standard expression for the excitation energy $U$
through the 
 saddle point $\beta^\ast=1/T$, i.e., $U=aT^2$.
Taking the preexponential Jacobian multiplier off the integral
 over $\beta$ in Eq.~(\ref{rhoE1F}) 
we substitute Eq.~(\ref{expbeta}) for $S(\beta) $
into Eq.~(\ref{rhoE1F}). 
Changing the integration variable $\beta$ to the new variable $z$,
$z^2=(-\partial^2 S/\partial \beta^2)^{\ast}(\beta-\beta^\ast)^2/2$,
and then calculating the error integral over $z$ by extending the
integration range to infinity, one obtains Eq.~(\ref{SPMgen}).
Here we used a general expression (\ref{Jac1F}) for the Jacobian
at the saddle point 
condition (\ref{expbeta}) for $\beta^\ast$. 
The critical quantity for these derivations is the ratio $\xi^\ast$;
see Eq.~(\ref{xiLIN}) for $\xi$ taken at $\beta=\beta^\ast$, $\xi=\xi^\ast$,
which is approximately proportional
to the semiclassical POT 
  energy shell correction, Eq.~(\ref{dedg}) (see Appendix \ref{appA}).

%

%-----------------------------------------------------------------------

\end{document}